%% file: main.tex
\RequirePackage{fix-cm}

\documentclass[smallextended]{svjour3}

\usepackage{amsmath,amssymb,amsfonts}
\usepackage{graphicx}
\usepackage{url}
\usepackage{xcolor, xparse}
\usepackage{subfig}
\usepackage{booktabs}
\usepackage{listings}
\usepackage{caption}
\usepackage{tcolorbox}

\usepackage{enumitem}
\usepackage{noindentafter}
\NoIndentAfterEnv{itemize}
\NoIndentAfterEnv{description}
\NoIndentAfterEnv{enumerate}
\setlist{
	leftmargin=14pt,
	itemsep=1pt,
}
\newcommand\mydescformat[1]{{\normalfont\itshape#1:}}
\setlist[description]{
    format=\mydescformat
}

\usepackage[noindentafter]{titlesec}
\titleformat{\paragraph}[runin]{\itshape}{}{}{}[.] 
\titlespacing{\paragraph}{0pt}{4pt plus 2pt minus 3pt}{4pt}

\newlength{\leftbarwidth}
\newlength{\leftbarsep}
\colorlet{leftbarcolor}{black}

\usepackage{framed}
\renewenvironment{leftbar}{%
    \MakeFramed {\advance \hsize -\width \FrameRestore}%
}{\endMakeFramed}

{\begin{leftbar}\itshape\noindent\hspace{-\leftbarwidth}\hspace{-1pt}
        \xspace}%
    {\end{leftbar}}%

{\begin{leftbar}\noindent}%
{\end{leftbar}}%

\usepackage{environ}
\NewEnviron{observationbox2}{\mybox{\BODY}}

\NewDocumentCommand\mybox{+m}%
{

\noindent\fbox{
\parbox{0.96\textwidth}{
#1
}}}%

\newcommand{\secref}[1]{Section~\ref{sec:#1}}
\newcommand{\tabref}[1]{Table~\ref{tab:#1}}

\newcommand{\figref}[1]{Figure~\ref{fig:#1}}

\newcommand{\RQ}[1]{RQ$_{#1}$}
\newcommand{\maven}[1]{\tool{Maven{#1}}} 
\newcommand{\gradle}[1]{\tool{Gradle{#1}}} 
\newcommand{\ant}[1]{\tool{Ant{#1}}}
\newcommand{\central}{\tool{Maven Central}}
\newcommand{\github}[1]{\tool{github{#1}}}
\newcommand{\java}{\tool{Java}}

\NewDocumentCommand\checkNum{+m}{{\color{red}#1}}
\RenewDocumentCommand\checkNum{+m}{#1}
\NewDocumentCommand\extension{+m}{{\color{blue}#1}}
\RenewDocumentCommand\extension{+m}{#1}

\RenewDocumentCommand\checkNum{+m}{#1} 
\NewDocumentCommand\code{+m}{{\small\ttfamily  #1}}
\NewDocumentCommand\tool{+m}{{\small\scshape  #1}}
\NewDocumentCommand\todo{+m}{}
\NewDocumentCommand\citeTodo{+m}{}
\newcommand{\ra}[1]{\renewcommand{\arraystretch}{#1}}

\begin{document}

\title{AROMA+: A Study of Factors Affecting Reproducible Builds in the Maven Ecosystem
\thanks{This is an extended version of ``AROMA: Automatic Reproduction of Maven Artifacts,'' accepted at FSE 2024. This manuscript is currently under review at Empirical Software Engineering (Springer).}
}
\subtitle{}

\titlerunning{Factors Affecting Reproducible Builds in Maven}

\author{Mehdi Keshani        \and
        Amirhossein Rahmati  \and
        Mohammad Hossein Aref \and
        Abbas Heydarnoori
}


\institute{Mehdi Keshani \at
              \email{mkeshan@bgsu.edu}           
           \and
                Amirhossein Rahmati \at
                \email{amir.hossein.rahmati.03@gmail.com}
           \and
                Mohammad Hossein Aref \at
                \email{m.h.aref2002@gmail.com}
            \and
                Abbas Heydarnoori \at
                \email{aheydar@bgsu.edu}
}

\renewcommand{\makeheadbox}{}
\date{}
    \maketitle

\begin{abstract}
Modern software engineering establishes software supply chains and relies on tools and libraries to improve productivity. However, reusing external software in a project presents a security risk when the source of the component is unknown or the consistency of a component cannot be verified. The SolarWinds attack serves as a popular example in which the injection of malicious code into a library affected thousands of customers and caused a loss of billions of dollars.
Reproducible builds present a mitigation strategy, as they can confirm the origin and consistency of reused components.
A large reproducibility community has formed for Debian, but the reproducibility of the \maven{} ecosystem, the backbone of the Java supply chain, remains understudied in comparison.
Reproducible Central is an initiative that curates a list of reproducible \maven{} libraries, but the list is limited and challenging to maintain due to manual efforts.
Our research aims to support these efforts in the \maven{} ecosystem through automation.
We investigate the feasibility of automatically finding the source code of a library from its \maven{} release and recovering information about the original release environment.
Our tool, \tool{AROMA+}, can obtain this critical information from the artifact and the source repository through several heuristics and we use the results for reproduction attempts of packages on \central{}.
Overall, our approach achieves an accuracy of up to \checkNum{99.8\%} when compared field-by-field to the existing manual approach. In some instances, we even detected flaws in the manually maintained list, such as broken repository links. We reveal that automatic reproducibility is feasible for \checkNum{32\%} of the packages on \central{} using \tool{AROMA+}, and \checkNum{12\%} of these packages are fully reproducible. We demonstrate our ability to successfully reproduce new packages and have contributed some of them to the Reproducible Central repository. Additionally, we highlight actionable insights, outline future work in this area, and make our dataset and tools publicly available. 


\end{abstract}
  
\section{Introduction}
\label{sec:introduction}


Software ecosystems, such as \maven{}, host libraries and build plugins in repositories, which form the backbone of many software supply chains. \central{} is the largest public repository for \tool{JVM}-based languages, allowing developers to reuse software and release their own packages there.

Relying on a supply chain has the risk of importing vulnerabilities into a project, and it either requires trust in the source or the ability to check the consistency of the packages. Unfortunately, research highlights an increasing number of attacks in recent years~\cite{lamb2021reproducible}. Some of them became popular, like the infamous SolarWinds hack in 2020, which resulted in the shipping of tainted binaries to thousands of customers, including the U.S. government, which was subsequently hacked~\cite{solarWindsAttack}. The \emph{Backstabber's Knife Collection} by Ohm et al.~\cite{ohm2020backstabber} documents 174 similar attacks.

It is obvious that trust alone is insufficient, and checks are necessary. Wheeler et al.~\cite{wheeler2005countering} were the first to propose a conceptual solution that involves comparing the compilation results of two different environments to detect modified binaries, such as those introduced in a malicious attack. A match is achieved when two software builds produce bit-for-bit identical binaries from the same source code, resulting in the same hash for the binary. If this can be accomplished, the binary is considered to be {\it reproducible}. 
This idea inspired the \emph{Reproducible Builds} initiative~\cite{reproducibleBuilds}, which recommends practices for improving software reproducibility. Many efforts, such as localizing unreproducibility~\cite{ren2018automated,ren2019root} or automatically addressing them~\cite{Ren2022Automated}, have been made.

So far, the community focus was \tool{Debian} and \tool{Linux}, which are currently \checkNum{$\sim$90\%} reproducible~\cite{DebainRepStats}, and reproducibility efforts for \maven{} packages are more limited.
{\it Reproducible Central}~(RC) represents a notable exception and provides tools for reproducing projects~\cite{reproducibleCentralRepo}.
The project also maintains a list of reproducible \maven{} projects, however, the reproducibility of the \maven{} ecosystem is still in its infancy. No official statistics exist, but the RC repository reports rebuild attempts for \checkNum{8,660} releases of 
\checkNum{914} projects. Compared to the \checkNum{10+M} releases of \checkNum{479K} unique \code{artifactId}s and \checkNum{64K} unique 
\code{groupId}s that exist on \central{},
it becomes clear that this invaluable effort only covers a fraction of packages. Furthermore, RC starts their reproducibility attempts at \github{} repositories, which do not necessarily equate to \maven{} releases. One such repository can have multiple sub-modules and releases. Our analyses also found links to \checkNum{64K} unique \github{} repositories on \central{}, which indicates that many repositories 
are currently missed.

Efforts to increase the reproducibility coverage are currently limited by several factors:
(i) reproduction is a manual, high-effort task, making it slow, unscalable, and prone to errors;
(ii) starting from source code repositories assumes that developers aim to create reproducible software, which is not the case for many unmaintained legacy projects;
(iii) current tooling often assumes the availability of the correct version of the source code for a given release. However, users who wish to reproduce a \maven{} package must first recover its source code; and
(iv) reproducibility is a binary metric and requires an \emph{exact}, \emph{bit-to-bit} match. Many factors influence this comparison, but not all differences are as critical as differences in the generated \code{.class} files. Even trivialities like file modification times or line endings can prevent a match. We believe that categorizing these cases as simply \emph{unreproducible} is too coarse-grained, and that we need to distinguish reproducibility on a \emph{spectrum}.

\smallskip
In this paper, we explore the possibility of automating the reproduction of large portions of the \maven{} ecosystem. We take the perspective of the community and try to achieve this goal without relying on the project maintainers. To achieve this goal, we investigate three key research questions:

\begin{itemize}
\item \RQ{1}: Can we recreate the link to the source code repository and commit of an artifact?
\item \RQ{2}: Can the original build environment be reconstructed after the fact?
\item \RQ{3}: To what extent is it possible to automatically reproduce \maven{} artifacts?
\end{itemize}

To answer these questions, we created a dataset of metadata for 480k packages. These packages represent one random version from every \maven{} project (\code{groupId:artifactId}).
This dataset lets us explore the factors that influence release practices and reproducibility on \central{} and we report on our empirical insights.

Our results are promising and show that an automated approach can recover up to \checkNum{$\sim$99.8\%} of the information manually maintained by RC.
We even identified inconsistencies in the manually maintained list, such as broken repository links. We successfully applied our approach in three practical use cases: reproducing packages that already existed on RC, identifying numerous reproducible libraries not previously listed by RC, and achieving near-perfect reproductions for binaries of some packages that did not aim to be reproducible. So far, we contributed \checkNum{three} new packages to RC's repository via \emph{pull requests}, all of which were accepted by their developers.

\smallskip
\noindent
This paper presents the following main contributions:

\begin{enumerate}
    \item A comprehensive study of the aspects that impact reproducibility in \central{}. 

    \item Heuristic approaches for recovering the repository link, release tag, \tool{JDK} version, \extension{release profiles, build tools,} and line endings for a given package.
    
    \item A tool for the Automated Reproduction of \maven{} Artifacts (\tool{AROMA+}).
    \extension{
    
    \item An evaluation of \tool{AROMA+} for the automatic reproduction of \maven{} and \gradle{} projects published on \central{}.
    
    \item Analysis of the impact of profiles on the reproducibility of \maven{} artifacts. }

\end{enumerate}

\smallskip

The remainder of this paper is structured as follows: We present our Experimental Setup in \secref{methodology}. In 
Sections~\ref{sec:rq1}, \ref{sec:rq2}, and \ref{sec:file-level-comparison}, we elaborate on our research questions, the methodologies that we used to answer them, and the results of our investigation. We then discuss the implications, future directions, and threats to the validity of our work in \secref{discussion}. \secref{related-work} presents the related work of this study. Finally, we summarize our work in \secref{conclusions}. All source code and data for this paper are available in a replication package~\cite{repPackage}.

\section{Experimental Setup}
\label{sec:methodology}

The primary focus of this study is the automated reproduction of \maven{} libraries to verify the consistency between source code and binaries. We investigate the feasibility of using only the source code and available metadata for a \maven{} package, without involving the project maintainers.
Throughout this study, we present empirical findings at each step for various reasons. Firstly, these findings justify our data-driven decisions in the methodology. Secondly, the data assists in setting defaults in projects like RC. For instance, by identifying the most popular JDK versions and using them as the default for reproducing packages, we can improve the likelihood of correct identification. Lastly, this approach enhances the transparency and verifiability of our results.

\subsection{Dataset Creation}
\begin{figure}
\centering
\includegraphics[ width=0.9\textwidth ]{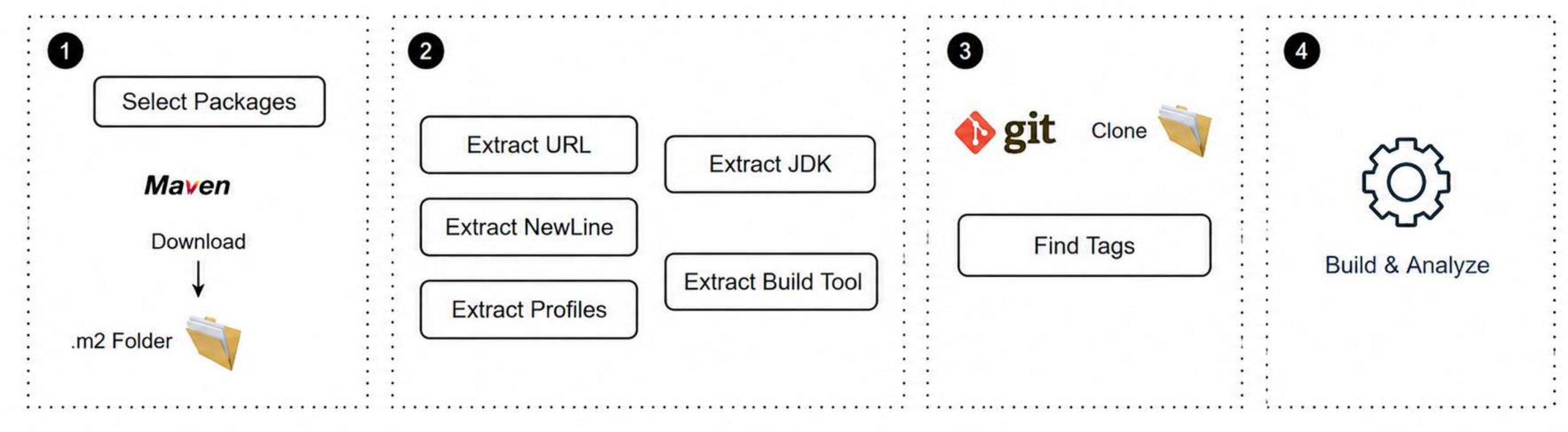}
\caption{Approach Overview 
}
\label{fig:overview}
\end{figure}
Our research requires data collection from both \maven{} and \github{}. \figref{overview} provides an overview of the methodology we employ for our data processing pipeline. \central{} publishes an index file weekly, listing all the released packages~\cite{mavenIndex}. We utilized the index files available up to \checkNum{May 17, 2023}. As of this date, \central{} contained \checkNum{10.3M} released versions of \checkNum{479,915} unique projects (\code{groupId:artifactId}). To ensure our dataset is representative of \central{} and to avoid bias towards projects with frequent releases, we sample one random release (version) for each unique project. We do not opt for selecting only the latest versions since our ultimate goal is to reproduce all of \maven{}, not just the new releases. This includes packages not designed for reproducibility, especially those predating this feature in \maven{}, thus requiring older versions. Furthermore, as not all projects use the latest versions and some neglect dependency updates, including the older versions becomes important.

We developed a scalable infrastructure for parallel data aggregation of selected releases. The scalability of this process is constrained only by available computational resources and network bandwidth. Leveraging a powerful server with \checkNum{256} cores, we completed the dataset population in \checkNum{3.5} hours. Building the projects required an additional \checkNum{8} hours. The infrastructure is extendable with custom analyzers, which can add supplementary information to the dataset. The system is deployed via \tool{docker-compose}, simplifying both the recreation of the dataset and periodic updates. The dataset creation uses the local \code{.m2} repository as a cache, downloading all \maven{} files to it, including the \code{pom.xml} files and archives. This creation process also populates a \tool{Postgres} database, which can be queried for insights about the ecosystem.

\begin{figure}
\centering
\includegraphics[width=\textwidth]{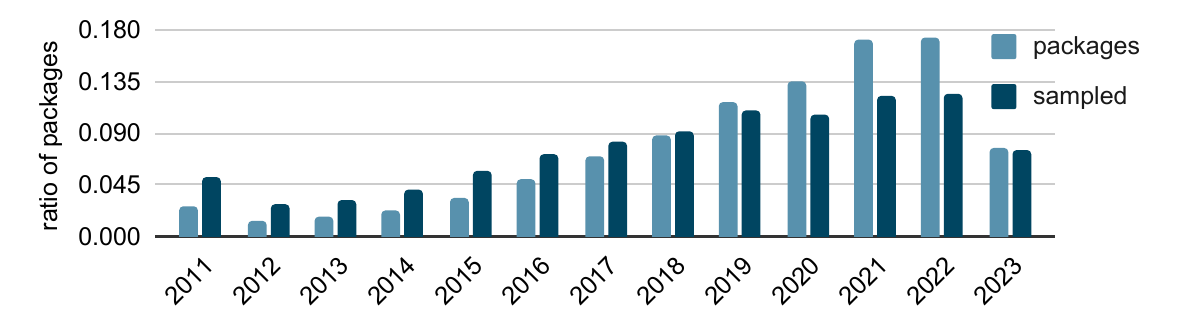}%
  \caption{Maven libraries and selected versions by release year.}
  \label{fig:maven_years}
\end{figure}

\paragraph{Statistics}
\figref{maven_years} displays the total number of releases on \central{} as well as the number of selected releases, grouped by their release year. This figure clearly demonstrates a significant growth in the number of libraries released over time. The plot indicates that our random sampling strategy yields a time-based distribution of releases that mirrors the overall distribution of \maven{} releases.
For our experiments, we successfully downloaded the majority of the selected packages, with only \checkNum{1.37\%} failing to download. A manual inspection revealed that most download issues were related to parent POMs not being hosted on \central{}. Our sample is detailed in \tabref{input-breakdown}. As this table shows, we successfully downloaded \checkNum{473,352} packages, out of which \checkNum{410,102} had archives and were included in our study.

\begin{table}
\centering
\caption{Studied dataset stats 
}
\label{tab:input-breakdown}
\input{tables/input-breakdown}
\end{table}

\subsection{Reproducibility}
\label{sec:reproducibility}
\extension{While the artifacts released on \central{} can, in principle, be built with any build tool, this study focuses on the most commonly used ones: \maven{} and \gradle{}.}. Reproducing a package presents numerous challenges. The most significant challenge is that build results depend on the build time. For instance, the current time may appear as the change time of the \code{.class} files or be included in embedded documentation. While adjusting the system time on the build machine can make smaller builds reproducible, this method is fragile. Even minor performance differences in the second build system can lead to varying checksums. \extension{Moreover, other sources of variability, such as file order, can also make a build unreproducible.}

To address these issues, the \maven{} build plugins support the \code{project.build.\-outputTimestamp} property for all time-related output and operations. This property can be specified in the POM file to fix a particular timestamp, facilitating reproducibility. Using this property addresses \emph{most} sources of variability in the default \maven{} plugins. However, even though the property can be set automatically during a build (e.g., through the \code{maven-version-plugin}~\cite{versionPlugin}), the vast majority of projects do not define it. In our dataset of \checkNum{$\sim$480K} projects, we could only find \checkNum{9,603} uses. Unfortunately, this property is not a golden bullet; other sources of variability also exist that can hinder successful reproduction.

\extension{In the case of \gradle{}, reproducibility can be improved by using the flags \code{preserveFileTimestamps} and \code{reproducibleFileOrder}. By default, \gradle{} artifacts may include original file modification times or store files in a nondeterministic order. Both issues lead to differing checksums across builds and make builds unreproducible. Setting \code{preserveFileTimestamps = false} removes the timestamp variability effect by ensuring that files in archives do not include their original modification times. Similarly, setting \code{reproducibleFileOrder = true} enforces a consistent ordering of files in the generated artifact. Together, these options address two major sources of non-determinism in \gradle{}'s build outputs, making artifacts more reproducible.}

\paragraph{Variability Sources}
\label{sec:method:rep-central}
The RC project stands at the forefront of reproducibility efforts in the \maven{} ecosystem~\cite{reproducibleCentralRepo}. The project has identified all sources of variability that typically exist in the build environment of \maven{} packages and has introduced a \code{.buildspec} format to formalize these build parameters for reproduction. A \code{.buildspec} requires the following information for reproduction:

\begin{itemize}
    \item A valid link to the repository and the exact tag used to mark the released commit.
    \item The \tool{JDK} version used for the compilation and build execution.
    \item The type of newline characters used on the build platform (e.g., \textbackslash{}n or \textbackslash{}r\textbackslash{}n).
    \item The build tool and the specific build command for the release.
\end{itemize}

The RC project offers advanced tools that utilize the \code{.buildspec} files to validate the reproducibility of a \maven{} build. Moreover, the project maintains a manually curated list of reproducible \maven{} packages and provides their corresponding \code{.buildspec} files. 

\paragraph{Reproduction}
Our study investigates the feasibility of automatically generating \code{.buildspec} files from the metadata that is available for \maven{} packages. The following sections introduce our heuristics that we devised to recover the required build environment of a release. After collecting all the information, we generate \code{.buildspec} files and compare our results with the manually curated ones provided by RC. We also use the RC tooling to attempt an actual reproduction of the release. RC employs \tool{Docker} containers to build projects, consistently resulting in the same output. We reuse their build scripts and do not modify their logic. This means that when comparing the local binary using RC scripts with the original binary on \maven{}, the comparison is effectively between the outputs of two different build environments: one from our build and one from the release on \maven{}, which the maintainers of the library previously built.

\section{\RQ{1}: Can We Recreate The Link To The Source Code Repository And Commit Of An Artifact?}
\label{sec:rq1}

\maven{} releases are often distributed with a source archive, however, these archives do not contain the required build files. As such, it is crucial to recover the source code from the original project repository, which can be linked in their POM.
However, this linking presents several challenges.

\paragraph{Non-Existing and Ambiguous Information}
Despite being mandatory information for \maven{} releases~\cite{mvnSCMRequirement}, not all projects link the source code repository in their POM. The links that do exist suffer from ambiguity, as there are multiple fields in which developers can declare the URL:

\begin{itemize}
\item \code{project.url}: Projects should link to their homepage, but many include the repository URL here.
\item \code{package.scm.url}: A public URL of the source code repository.
\item \code{package.scm.connection}: A link to a read-only version of the source code repository.
\item \code{package.scm.developerConnection}: Similar to \code{package.scm.connection}, but with write access.
\end{itemize}

Despite the clear recommendation to specify the repository URL in the \code{package\-.scm.url} field, in practice, all fields are used interchangeably. The \maven{} manual defines the format for each field~\cite{maven-pom}, e.g., the \code{package.scm.connection} should be \code{scm:<provider>:<provider-specific>}. However, this format is not enforced and is not strictly adhered to by many.

\paragraph{Validating Links}
We need a valid repository link to rebuild releases from their sources. Therefore, we verify the provided repository links, similar to previous work that studied the prevalence of broken links~\cite{Liu2021,akhavan2025linkanchor,Rostami2021}. This step is not only crucial as a starting point for our work but it also provides first insights into linking practices and popular hosting platforms.

Our approach is limited to \tool{Git}, currently the most prevalent version control system~\cite{gitMostPopular}. We leverage its \code{ls-remote} command, which requests references to branches and tags from a remote repository without the need to clone or fetch from it first. Successful execution conclusively demonstrates the validity and accessibility of the queried repository.
We validate all four fields separately, storing the results. Links failing this validation are either invalid or associated with different version control systems.

\paragraph{Release Tags}
After identifying the repositories, we need to pinpoint the exact commit that has been released for successful reproduction. Using release tags~\cite{gitTatting} is a common strategy to mark releases in a repository. In \tool{Git}, tags are unique markers for distinct repository states. They can be used to label pivotal milestones, like major versions or feature releases. As such, we inspect the tags in the source code repository and try to find tags that correspond with the version of the \maven{} release. Unfortunately, the format is not unified, and a wide range of tagging schemes exists in practice. For instance, a release of version \code{1.0.0} could be tagged with \code{v1.0.0}.

\subsection{Where Do Maven Projects Host Their Source Code?}

The established repository links allow us to investigate which hosting providers are commonly used for the source code repositories. Combined with the timestamp of each release, we can also analyze an evolution throughout the history of the ecosystem. Such insights highlight the interdependency between \maven{}, which serves as a binary ecosystem, and the code repositories that function as source code ecosystems.

\paragraph{Methodology}
To examine the market share evolution of each repository host, we parse each URL field, extract the hostname, and group the hostnames by year. For each year, we count the number of repositories for each service and determine the percentage market share of each hosting service by calculating the ratio of its count with the total number of repositories found in that year.

\begin{figure*}
    \centering
    \includegraphics[width=.9\textwidth]{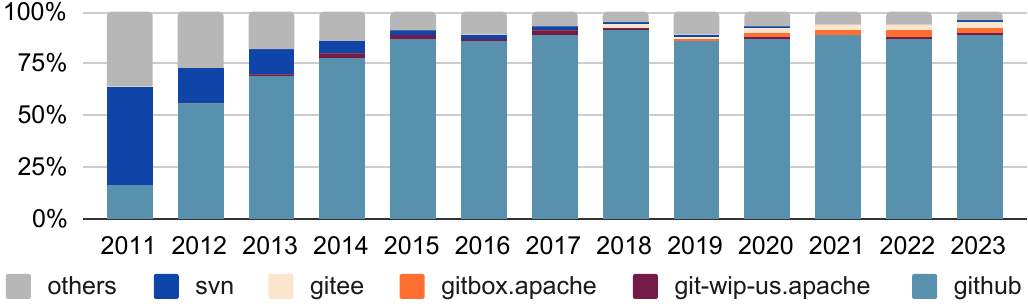}
    \caption{Market share of repository hosts per year for package.scm.url field.}
    \label{fig:market_share}
\end{figure*}

\paragraph{Results}
Overall, \checkNum{83\%} of packages with \code{scm.url} utilize \github{}.
\figref{market_share} depicts the market share of all source code repositories over time for this field. Moreover, since other URL fields exhibit similar patterns, our focus here is exclusively on the results pertaining to the \code{scm.url} field. Given the abundance of repository hosting services, we included only the five most popular ones and grouped the remaining ones into the \emph{others} category. 
As illustrated in \figref{market_share}, \github{} experienced rapid growth from \checkNum{2011} to \checkNum{2014}. During this period, other repositories, especially \tool{SVN}, maintained some degree of popularity. However, they were mostly replaced by \github{} in recent years. By \checkNum{2015}, \github{} had secured a market share consistently exceeding \checkNum{85\%}. In addition to \github{}'s dominance, an emerging trend indicates a rising preference for alternative repository hosts. Among them, \emph{gitee}, \emph{git-wip-us.apache}, and \emph{gitbox.apache} are particularly noteworthy. However, these repositories are mostly exclusive to Apache projects and are often mirrored on \github{}. The figure also captures Apache's shift from the \texttt{git-wip-us.apache.org} domain to \texttt{gitbox.apache.org} in late \checkNum{2018}~\cite{apache-change-repo}.

\subsection{How Reliable Are The Source Code Repository Links on Maven?}
\label{sec:results_url}

It is not enough to extract a seemingly valid URL, it is important for our evaluation that we ensure that these links point to the correct repository. First, if we build a different repository than RC does, the results of our builds will naturally differ, making them incomparable. Second, discovering new \maven{} reproducible packages is of little value if we are building the incorrect repository.

\paragraph{Methodology}
After obtaining a valid repository link for each package, we compare them to the repository links of RC \code{.buildspec}s to assess the accuracy of our approach. Since RC is mainly created manually, it serves as a good benchmark for evaluating our method.
The manually curated \code{.buildspec} files are only \emph{one possible way} to reproduce a release; other \code{.buildspecs} could lead to the same result.
Using these files in our evaluation gives us a lower bound, and even a negative match might lead to a reproducible build in practice.

We recognize that several URLs in a \code{.buildspec} point to mirror repositories and a direct link comparison would result in many mismatches.
As such, we decided to rely on the command \code{git ls-remote url HEAD} for comparison, which returns the commit hash of the default branch. If the hashes of both URLs match for a particular package, we consider the repositories to be identical. 

\paragraph{Results}
\checkNum{Out of the 473,352 packages that we downloaded,  442,360 (representing 97.34\% of packages with at least one \code{url})} provided a repository URL in the \code{scm.url} field. This high percentage is attributable to the mandatory requirements of \maven{} libraries~\cite{maven-requirements} specified by the ecosystem. Nevertheless, not all developers utilize the \code{scm.url} field as intended. Some include a URL pointing to the project's homepage rather than the source code repository, which should be specified in the \code{url} field instead. Conversely, there are instances where developers either place the source code repository URL in both fields or solely in the \code{url} field. Some even provide the SSH URL for the repository, which might require authentication. However, converting such a link to an \code{HTTPS} URL makes it accessible. We automated this conversion process.

\tabref{urls} shows the results of our URL validation. The left part lists the percentage of packages that contain each field, while the right part breaks this number down into the percentage of packages for which the identified link is valid.
Notably, a significant percentage \checkNum{(74.09\%)} of links in the \code{scm.url} field were validated, while a considerably lower percentage \checkNum{(37.7\%)} in the \code{url} field met the same criterion. Since the \code{url} field is intended for the project's homepage, this difference is expected; after all, a webpage cannot be verified as a repository. However, the presence of \checkNum{37.7\%} valid repository links in the \code{url} field also indicates that many projects use their repository as their homepage.

Overall, \checkNum{79.2\%} of the packages have at least one verifiable repository link. Of the packages with invalid links, \checkNum{6.87\%} contain URLs that are unparseable and could not be validated. These cases include malformed links, empty strings, and strings that are not URLs.

\begin{table}
\centering
\caption{Percentages of URL field usage and valid URLs for each}
\label{tab:urls}
\resizebox{\textwidth}{!}{%
\input{tables/url-field-usage}
}
\end{table}
In total, our dataset contains \checkNum{1,517} packages for which RC offers a \code{buildspec}. \checkNum{97.5\%} of the URLs we identified align with the repositories that RC also references. Through manual inspection of the \checkNum{2.5\%} non-matching cases, we determined two categories of mismatch:
\begin{itemize}
    \item Instances where RC provided a URL requiring authentication for access, while the URL we identified is public and matches the package.
    \item Cases in which RC offers a link that results in a \code{404} error, but our URL points to a relevant public repository corresponding to the package.

\end{itemize}

Most importantly, we did not encounter any instances where our method provided a link that contradicted the information in RC.

\subsection{Can We Find The Corresponding Version Tags in Source Code Repositories?}

Developers of a project can leverage release tags to roll back to a specific version for tasks such as debugging and bug fixing. Yet, from \maven{}'s perspective, tags remain hidden. If the conventions developers use for \maven{} releases differ from the tags in their codebase, then only a developer's historical knowledge can bridge this gap. In this research question, our objective is to automate the mapping between \maven{} releases and \tool{Git} tags. 

\paragraph{Methodology}
\label{sec:tags-method}
To discover more version mappings, we do not limit our tag search to exact matches and inspect all tags used in the validated repositories to find recurring tagging practices. From this list, one author has identified patterns, calculated their frequency, and removed their matches until no other repeating patterns existed. We verified each pattern by ensuring that a second author cannot find mismatches in \checkNum{10} randomly selected matches for each pattern. Overall, we identified \checkNum{45} common tagging patterns, and \checkNum{our replication package~\cite{repPackage} contains the full catalog}.

\newcommand{\VER}{{\it \texttt{<<version>>}}}
\newcommand{\NUM}{{\it \texttt{<<number>>}}}

\paragraph{Results}
From a total of \checkNum{354,948} packages that have at least one valid URL, we successfully identified \checkNum{234,674 (66.1\%)} tags using the tagging patterns. Packages that did not align with our patterns either lacked version tags or adhered to a specific convention unique to their project, which differed from those we identified.

\tabref{tags} presents the prevalence of our top 10 tagging patterns and also showcases examples of tags used in \maven{} libraries.
The most common pattern shown in the table is \code{v\VER}, which indicates that the package version that can be derived from the \maven{} coordinate is prefixed with the character \emph{v}, leading to tags such as \emph{v1.0.0}.
Notably, this pattern is present in \checkNum{113,243} packages within our dataset. Additionally, some packages incorporate their \code{artifactId} or parts of it into their tag names. An \code{artifactId} might consist of several parts, each separated by a dash (\code{-}). We label these parts as \code{p\NUM}. For illustration, the package \code{com.adobe:aio-lib-java-ims:0.0.4} contains \checkNum{four} parts (p1: \code{aio}, p2: \code{lib}, p3: \code{java}, and p4: \code{ims}).
Within our dataset, \checkNum{639 (0.3\%)} packages follow the pattern \code{p1-p2-p3-\VER}. The first \checkNum{two} patterns listed in the table account for \checkNum{83\%} of the cases we identified.
Our results suggest that the vast majority of \maven{} libraries either use the version number as a tag or prepend the character \emph{v}.

\begin{table}
\centering
\caption{Top 10 tagging patterns}
\label{tab:tags}
\resizebox{\textwidth}{!}{%
\input{tables/top10-tagging-patterns}
}
\end{table}

Overall, in \checkNum{93.4\%} of the cases, the tags that we identified are identical to the ones RC suggests. We manually inspected the remaining \checkNum{6.6\%} of tags and found two categories of mismatching tags:

\paragraph{Commit Hashes} RC references a commit hash instead of a tag label. In some cases, it is the same commit that our identified tag points to. Occasionally, when a branch was merged during the release, RC includes the parent or child commit of our identified tag. In these cases, merge commits were used solely for adjusting the versions, so the difference between RC and our approach is negligible. We found two instances where RC specified a seemingly unrelated commit. In the first instance, a comment in the \code{.buildspec} notes that the binaries on \central{} were not built from the tag, which suggests an oversight or poor practice by the library maintainers. In the other instance, the \maven{} plugin version has changed in the commit of RC. We speculate that the package developers attempted to update a plugin, which made this commit unreproducible. RC might have realized this and instead decided to point to the correct commit instead of the tagged one. However, without better documentation or practices for such cases, it is challenging for others to interpret or automate them.
\paragraph{Tag Names} RC pointed to a different tag name than what our method identified. In some cases, repositories featured identical releases but with different tag names; one was detected by our method, and the other was found by RC. When comparing the content of both releases, they were identical. Thus, our approach is essentially the same for such cases as RC. In one instance, RC pointed to a tag that did not exist in the repository. However, a branch with the same name was present, enabling a checkout and build. Nevertheless, the tag we identified was also accurate since it pointed to the last commit of the same branch. In another instance, we pinpointed a tag as \code{1.0.0}, while RC referred to \code{org.apache.felix.feature-1.0.0}. We noted in the repository of this package that developers versioned each sub-module individually. Thus, \code{1.0.0} represents the version \code{1.0.0} of the entire project, whereas \code{org.apache.felix.feature-1.0.0} pertains to the release of the particular sub-module we were looking for. In this particular example, our approach may not be able to find the accurate release; however, this pattern of versioning is not advised. Consistent versioning across the entire project is the recommended practice. We observed that such cases are exceptions and do not occur frequently.

In conclusion, our observations did not reveal any intrinsic flaws in our approach to identifying tags. However, during the manual inspection, we found that there is a lack of standard practices and procedures when it comes to the reproducibility of packages and releases in general, which makes it even more challenging to automate.

\begin{tcolorbox}
\RQ{1}: Can we recreate the link to the source code repository and commit of an artifact? Yes, for a total of 354,948 packages, we identified at least one valid repository URL. Among these, we successfully pinpointed commit tags for 234,674 using our identified tagging patterns. In summary, for 57.2\% of the packages, we can recreate both their link and the commit associated with the artifact.
\end{tcolorbox}


\section{\RQ{2}: Can The Original Build Environment Be Reconstructed After The Fact?}
\label{sec:rq2}

After obtaining the source code, a reproduction attempt requires building the package. Additional details about the original build environment need to be defined in the \code{.buildspec} to use the RC tooling for the reproduction. We have devised several heuristics to recover these details from the source code or the released artifacts.

\paragraph{JDK}
\label{sec:method:jdk}
\java{} is the primary language for \tool{JVM}-based programs that are released on \maven{} and the development of \java{} programs requires a \emph{Java Development Kit}~(JDK). The chosen \java{} version notably affects the development and release process of libraries. New \java{} versions introduce new features, refine existing ones, fix bugs, and address vulnerabilities. As the compiler changes with each version, it also has an important effect on a reproduction attempt for a library from its source code, as the usage of different \java{} versions for compilation can result in different bytecode.

To collect data about the \tool{JDK} version used to originally build the libraries, we employ two heuristics. The first heuristic focuses on the archive's metadata file \code{META-INF/MANIFEST.MF}.
If present, this file usually provides information about the \tool{JDK} version in the \code{Build-Jdk-Spec} field; older versions of the \maven{} archiver~\cite{jdkManifestReplaceJira,jdkManifestReplaceGithub} have used the \code{Build-Jdk} field instead. The \code{Build-Jdk} field typically contains detailed version data, including the minor version. For the purpose of rebuilding the packages, we only need the major versions. As such, we extract only that major part. For instance, from the version \code{11.0.10}, we only retrieve \code{11}. It is worth noting, however, that a significant number of archives lack these entries. In such cases, we fall back to the configured compiler version, which can help reduce the \emph{search space} of \tool{JDK} candidates that could have been used for the release.

\paragraph{Compiler Configuration}

The source code compilation of \maven{} projects is configured within the POM through the \emph{Apache \maven{} Compiler Plugin}~\cite{mavenPlugin}.
While third-party plugins might also support compilation, we focus solely on the \emph{Apache \maven{} Compiler Plugin} as the main and default compilation tool.
It allows specifying the expected version of the \java{} \code{source} and the \code{target} version of the generated bytecode. Knowing about these versions eliminates \tool{JDK} versions that precede these releases as viable candidates, as only newer \tool{JDK}s can compile to old versions, but not vice versa. We use the extracted versions as the lower bound for the \tool{JDK} version and build the project using this specific version and all newer releases with \emph{Long Term Support} (e.g., Java 11, 13, 17) that were available at the release date of the library. With the assistance of the \emph{file-level comparison} (See \secref{file-level-comparison}), we can then identify the best version that maximizes reproducibility. In cases where we cannot find any version for the package, we try all \emph{LTS} versions.

To streamline the information extraction process from the POM, we utilize the built-in model of \maven{}~\cite{mavenModel}. This model offers a standard structure for accessing and extracting the required data from the POM. It is important to understand that POM supports numerous features for managing and inheriting properties among different modules. For instance, projects without actual \java{} code can still possess compiler configurations. These projects are typically shipped as \emph{pom} packages. They act as containers, and sub-modules can inherit configurations and dependencies from them. In this study, we consider such inheritance from the project's parent, configurations of plugins, and \emph{pluginManagement}~\cite{pluginManagement}. We also account for properties and \maven{} default values.

\paragraph{Line Ending}
Line endings or newline characters denote the end of a line in a text file and the beginning of a new one. These characters play an important role in formatting and displaying text across various platforms and editors. The two common line-ending conventions are \code{lf} (Line Feed) and \code{crlf} (Carriage Return plus Line Feed). \code{lf} is primarily used in \tool{Unix}-based systems, including \tool{macOS} and \tool{Linux}, and is represented as \code{\textbackslash n}. Moreover, \code{crlf}, which is common in \tool{Windows} environments, is represented as \code{\textbackslash r\textbackslash n}. Inconsistencies in line endings can result in incorrectly displayed text or errors in certain contexts. Line endings can also impact reproducibility, especially in build-generated files like \code{MANIFEST.MF} and \code{pom.properties}.

Providing the correct line endings used during the package release in the \code{.buildspec} is essential, allowing the RC tooling to eliminate variations caused by varying line endings. If the newline is \code{crlf}, the argument \code{-Dline.separator=\-\$'\textbackslash r\textbackslash n'} is applied when building the project. Additionally, during the git repository fetch, all newlines are converted to \code{crlf} using the \textit{unix2dos} tool.

To identify the correct line ending, we cannot examine source files from the repository, as committed files were created by a developer on a separate computer. Additionally, files can also get altered by \tool{Git} on checkout. Instead, we rely on the \code{pom.properties} file, which is auto-generated 
during the build and then added as metadata to the artifacts. This file presents a reliable source as its line endings reflect the environment in which \maven{} built the original release. In cases where we cannot extract line endings from the file, we follow a trial and error approach and use both \code{lf} and \code{crlf} to build the package. If one fails, we can repeat the reproduction with the alternative.

\paragraph{Build Tool}

Build tools are fundamental to modern software development, as they automate repetitive and error-prone tasks such as compiling source code, running tests, managing dependencies, and packaging applications for release. Among \java{} projects, \maven{}, \gradle{}, and \ant{} are considered the three most widely used build tools. \maven{} is often preferred for enterprise projects. \gradle{} is popular in modern projects, such as Android development. \ant{}, while older, is still in use for legacy systems and highly customized build processes. Identifying which build tool a project uses is an important step toward automatically reproducing its artifacts.

To identify the original build tool used to release a package, we search for evidence in both the artifact hosted on \maven{} and its corresponding \tool{Git} repository. We begin by examining the artifact's metadata for any information that may reveal the build tool used during its creation. For \maven{}, we inspect the \code{pom.properties} file, which is automatically generated during the build process. This file often includes a comment such as \code{\#Generated by Maven}, providing strong evidence that the artifact was built with \maven{}.
For \gradle{} and \ant{}, we analyze the manifest file for entries that indicate the use of these tools. Specifically, we examine the main attributes section for references to the build system. If an attribute corresponding to \ant{} is present, such as \code{Ant-Version}, the artifact is classified accordingly. A similar approach is applied to \gradle{}, where we look for attributes such as \code{Created-By}, which often indicate that the artifact was built with \gradle{}.

When developing a project, each build tool requires a specific configuration file that defines essential build parameters such as artifact versions, and dependencies. \maven{} projects include a \code{pom.xml} file; \gradle{} projects use \code{build.gradle}, \code{settings.gradle}, or their \tool{Kotlin}-based equivalents (\code{build.gradle.kts}, \code{settings.\-gradle.kts}); and \ant{} projects use a \code{build.xml} file. If searching the metadata does not yield any conclusion, we proceed by inspecting the contents of the artifact itself (e.g., \code{war}, \code{jar}, etc.) for the presence of these configuration files. We infer the build tool as follows: the presence of a \code{pom.xml} file indicates a \maven{} build; any \gradle{}-related files (e.g., \code{build.gradle}, \code{settings.gradle}, \code{build.gradle.kts}, \code{gradlew}, \code{gradlew.bat}, or \code{settings.gradle.kts}) suggest the use of \gradle{}; and a \code{build.xml} file indicates a build with \ant{}.

In cases where metadata extraction does not yield conclusive results, we attempt to identify the build tool by analyzing the structure of the artifact's source code repository. Many open-source projects include build configuration files in their version control repositories, which serve as an indicator of the build tool used. We begin by retrieving the repository's contents, and then search for specific build configuration files. The presence of particular filenames uniquely identifies the build tool: As before, \code{build.gradle} or \code{settings.gradle} indicate \gradle{}, \code{build.xml} marks \ant{}, and the presence of \code{pom.xml} classifies the project as \maven{}.

\paragraph{Build Command}
\label{sec:command}
The configuration-based design of \maven{} makes it possible for the actual build command used to create a release to be reduced to basic instructions. While the details can differ for each project, such as relying on release profiles or certain environment variables, RC has found a meaningful default command that works across a wide range of projects. Their recommendation is to activate the \code{-DskipTests}, \code{-Dmaven.javadoc.skip}, and \code{-Dgpg.skip} flags to skip test execution, the generation of \emph{javadoc}, and the \code{GPG} signing process. None of these three steps affect the generated binaries; therefore, they can be safely skipped. Notably, the signatures are stored in separate files, and without knowing the private keys, reproduction would be impossible.

For our reproduction attempts, we adjust this default command of RC with further options. Since our recovered repositories potentially contain multi-module projects, we want to avoid building the whole repository every time, which would slow the build and increase the risk of a build failure. Therefore, we add the \code{-pl :{\it \texttt{<artId>}}} parameter, which limits the build to only the specific artifact that we are trying to reproduce. We also include the \code{-am}/\code{--also-make} flag to ensure that dependencies to other modules in the multi-module projects get built as well; otherwise, the build would fail. Ultimately, we use the command \code{mvn clean package -pl :{\it \texttt{<artId>}} -am -DskipTests -Dmaven.javadoc.skip -Dgp\-g.skip} to build the packages for reproduction. \extension{In addition to these flags and attributes, RC occasionally specifies a build profile for some projects in its build command. Since build profiles are project-specific, we investigate their automation and impact on reproducibility separately.}

\extension{
\paragraph{Build Profiles}
\maven{} build profiles provide a mechanism for defining alternative build configurations within a project's POM file. They allow developers to tailor the build process for specific scenarios, such as different operating systems, \tool{JDK} versions, deployment targets, or packaging preferences. Profiles are defined in the POM files under the \code{<profiles>} section and are identified by a unique \code{<id>}. Build profiles can significantly affect the build output, as they may alter various POM elements such as dependencies, plugins, compiler settings, and resource configurations. Our goal is to extract and analyze \maven{} build profiles data to assess their impact on build reproducibility. 

There are multiple ways to activate \maven{} build profiles. One method is manual activation, where users explicitly specify the profile using the \code{-P} flag during the execution of the build command. In addition to manual activation, some profiles are configured with an \code{<activation>} block in the POM file that defines conditions under which they should be triggered automatically. These conditions can be based on the operating system \code{<os>}, the \tool{JDK} version \code{<jdk>}, the value or presence of a specific system property \code{<property>}, or the existence or absence of certain files \code{<file>}. Additionally, a profile can be marked as active by default by setting \code{<activeByDefault>true</activeByDefault>} in its \code{<activation>} block
~\cite{mavenProfiles}. We extract profile information from these sections. Moreover, some build-related plugins, such as the \maven{} Release Plugin~\cite{mavenReleasePlugin}, may declare specific profiles to be enabled during their execution. As such,  \maven{} Release plugin can be a reliable source for determining which profiles were applied during a release.
Therefore, we also extract data from this Plugin.
Specifically, we analyze the plugin's \code{<configuration>} block to identify profiles declared under the \code{<releaseProfiles>} element. These profiles are activated during the release process and are likely to affect the reproducibility of build outputs.

}

\subsection{Which JDKs Were Used to Build Maven Libraries?}

Finding the correct \tool{JDK} is not only one of the most fundamental steps in our reproduction efforts, but it also allows us to study the distribution in the ecosystem. This can, for example, help library maintainers choose versions that result in more compatibility, as well as help RC maintainers try the most prevalent versions as their default \tool{JDK}.

\paragraph{Methodology}
To find the build \tool{JDK} versions, we analyze the \code{MANIFEST.MF} files of the packages in our dataset.
Out of \checkNum{410,102} packages with archives, \checkNum{204,394 (50\%)} defined the \code{Build-Jdk} field, and \checkNum{41,809 (10\%)} defined the \code{Build-Jdk-Spec} field in their manifest files. In this section, we report the statistics of analyzing the \code{Build-Jdk} field as more packages define this field.

\begin{figure*}[t]
    \centering
    \includegraphics[width=0.7\textwidth]{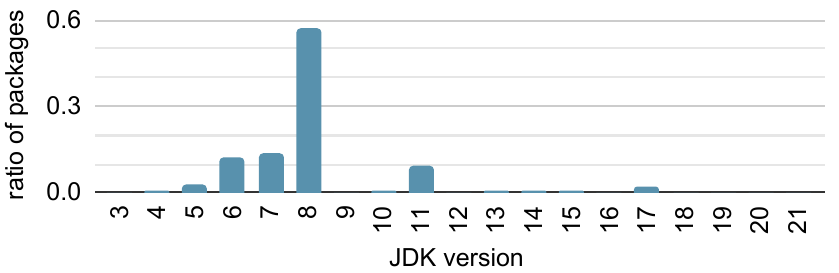}
    \caption{JDK versions used by Maven libraries.}
    \label{fig:jdk-versions}
\end{figure*}

\paragraph{Results}
The predominant \java{} version utilized is \checkNum{\java{} 8}, accounting for a significant \checkNum{57\% (116,607)} of the packages. When considering the \emph{Long-Term Support} (LTS) versions, namely \checkNum{\java{} 8, 11, 17, and 21}, they collectively make up about \checkNum{68.5\%} of the archives.
As the newest \tool{JDK} version (\java{} 20) was released only \checkNum{two} months before this study was conducted, it is not surprising that we could only find \checkNum{52} packages that use it.
It is noteworthy to mention that older non-LTS versions, specifically \java{} 5, 6, and 7, have a considerably higher presence compared to more recent non-LTS versions. A comprehensive breakdown by major version can be observed in \figref{jdk-versions}. Only one artifact, specifically~\cite{jdk-21}, appears to be compiled in \checkNum{Java 21}. Given that \checkNum{\java{} 21} has not been released at the time of writing this article, it is likely that this artifact is built using an early-access version.

\begin{figure*}[t]
    \centering
    \includegraphics[width=.9\textwidth]{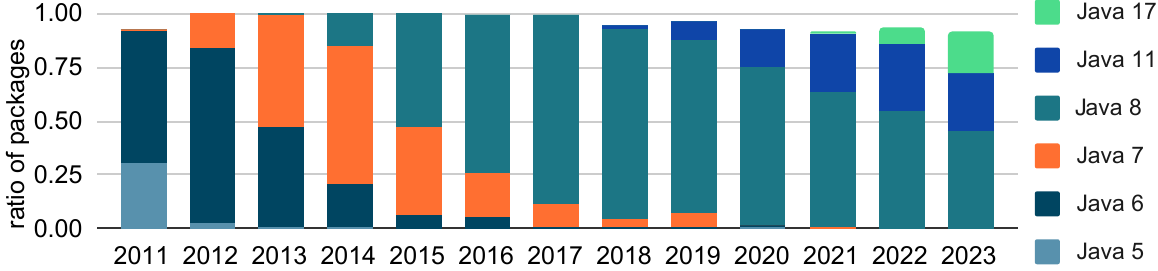}
    \caption{JDK versions used by Maven libraries over the years. 
    }
    \label{fig:jdk-versions-years}
\end{figure*}

\figref{jdk-versions-years} illustrates the distribution of \tool{JDK} versions per year. This figure includes only the top \checkNum{six} versions, comprising three LTS versions (\checkNum{8, 11, 17}) and three non-LTS versions (\checkNum{5, 6, 7}). It is important to note that other versions are excluded from this figure, which is why some bars have values less than one. The figure highlights the relatively rapid adoption of LTS versions (\java{} 8, 11, and 17). These versions become among the most popular choices within two years of their respective releases. For instance, \java{} 8 was released in \checkNum{2014} and became the most popular version by \checkNum{2016} with a share of \checkNum{26.6\%}. Interestingly, \java{} 8 maintains its dominance in the ecosystem even after newer LTS versions are introduced. The figure also highlights a pattern of quick growth followed by a prolonged decline for non-LTS versions. However, this pattern changes after the introduction of the first LTS version, \java{} 8. Its dominance in the ecosystem persists much longer than the popularity of preceding versions.

Our analyses have revealed that the artifact compiled in early-access \java{} 21 is not a unique case; we found several other instances of compilation in previous early-access \java{} versions as well.
We can also still find new releases in \java{} versions that are no longer officially supported. This trend appears in approximately \checkNum{21K} packages. It is important to note that this is only an estimation due to the limited information available regarding end-of-support dates for older versions. Furthermore, if these dates are available, they can vary by multiple years depending on the \java{} vendor.

The next method to recover the \tool{JDK} version involves examining the compiler configuration, as discussed in \secref{method:jdk}. In many instances, the \maven{} compiler plugin is not configured, accounting for \checkNum{45.02\% (213,106)} of the packages. However, it is configured in \checkNum{260,246} packages. When the \maven{} compiler plugin is utilized, both the \code{source} and \code{target} versions are specified in \checkNum{95.89\% (249,542)} of the cases. On the other hand, neither one is specified in \checkNum{3.92\% (10,202)} of the packages. Interestingly, even though it is designed to allow different source and target versions, in almost all cases where both are present, they match. Only a tiny fraction (\checkNum{0.15\%/385}) specify different versions. Although this feature serves its intended purpose, it is rarely used. Exclusively specifying just one of the two versions is also rare: a mere \checkNum{376 (0.14\%)} packages specify only the source version, while \checkNum{126 (0.05\%)} packages utilize only the target version.
\checkNum{64\%} of packages that specified both \tool{JDK} and source versions use the same version for both. However, \checkNum{35\%} use a newer \tool{JDK} for building than the one specified for the source. In rare cases (0.0012\%), there are even source fields that are greater than the compiling \tool{JDK}. These cases might be mistakes in the manifest files since older versions cannot compile more modern source code.

When compared to RC, our identified \tool{JDK} version matched in \checkNum{99.5\%} of the cases, and we manually inspected the very few differences. Determining the exact version that library maintainers used to build the released library is challenging since none of these projects have a manifest file. RC also does not provide details on how they arrived at the versions used in these packages. However, we discovered \github{} workflow files in the corresponding \github{} releases of these packages, and the \java{} version used in their \tool{CI} matched one of the LTS versions that we had identified.
In contrast, RC assigned non-LTS versions to these packages, which we could not verify within these projects.

\subsection{What Line Endings Do Maven Libraries Use?}

For the reuse of a package, it is irrelevant on which platform it has been built. However, understanding the distribution of build platforms can guide future research on reproducibility.

\paragraph{Methodology}
As explained in the introduction to this section, we extract the information about line-endings from the \code{pom.properties} file.
We calculate basic statistics about the file endings and group the results by year.

\paragraph{Results}
Overall, \checkNum{63\% (258,809)} of the \maven{} libraries with archives possess the \code{pom.properties} file.
Interestingly, \checkNum{212,340 (82\%)} of the packages employ \code{lf} line endings, while only \checkNum{46,469 (18\%)} use \code{crlf}. A historical analysis of the distribution throughout the history of \maven{} has revealed a consistent ratio of roughly \checkNum{80/20} in favor of \code{lf} over the years.

Our identified line ending aligns with RC's results in \checkNum{98\%} of the cases. In the few instances in which the results differed, RC employs \code{crlf-nogit} and \code{no-auto.crlf} as line endings, while we utilize \code{crlf}. We were unable to determine the reason RC adopts these line endings, as our search through their documentation and code yielded no answers, but we assume that this hints at the way \tool{Git} is (not) supposed to take care of the line endings~\cite{gitNewline}. Upon checking the \code{pom.properties} of those packages, we found that the releases used \code{crlf}. Thus, we are confident about our results.

\extension{
\subsection{How Do Maven Libraries Use Build Profiles?}
\label{sec:RQ2-profiles}

Understanding common usage patterns of build profiles is helpful for reconstructing build environments, as they can greatly impact the build outcome. Moreover, future work on reproducibility can benefit from such knowledge.

\paragraph{Methodology}
To better understand how profiles are used across the \maven{} ecosystem, we analyze their prevalence, origin (own versus inherited), and activation patterns. 
As explained in the introduction to this section, we extract build profile
information from POM files. We use the built-in \maven{} model~\cite{mavenModel}, recursively resolving parent POMs to account for profile inheritance. Profiles are collected and categorized based on their origin and level in the inheritance chain.

For each identified profile, we extract key attributes, including its \code{id} and \code{<activation>} block. We analyze the activation block to determine whether the profile was activated by default or conditionally activated based on criteria such as JDK version (\code{<jdk>}) or operating system characteristics (\code{<os>}).
To identify the profiles used during releases on \central{}, we extract those triggered by the \maven{} Release Plugin and refer to them as release-activated profiles.

Build profile is not a standalone field in the \code{.buildspec} and is occasionally included in the \code{mvn} section of build commands. Therefore, to compare our extracted profiles with those specified by RC, we parse RC's build command and search for a \code{-P} flag. We then compare the profiles indicated by these \code{-P} flags with the profiles we extracted from the \maven{} Release Plugin.

\paragraph{Results}

\begin{table}
\centering
\caption{Distribution of Maven Packages by Profile Presence}
\label{tab:profile-presence}
\input{tables/profile-presence}
\end{table}

\begin{figure*}
    \centering
    \includegraphics[width=.9\textwidth]{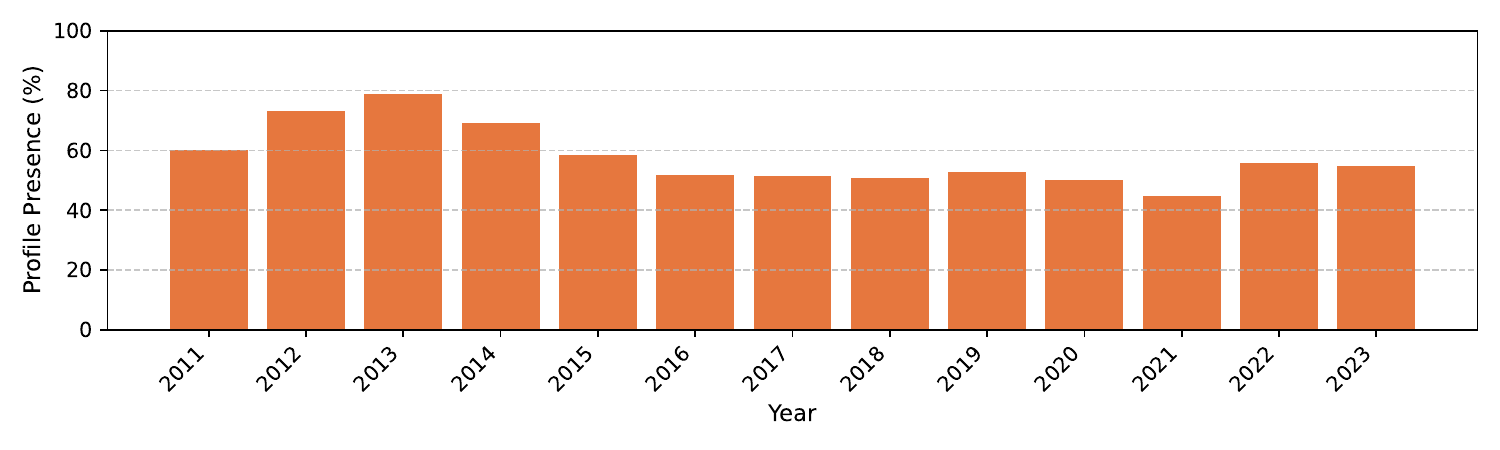}
    \caption{Profile presence in Maven packages over the years. 
    }
    \label{fig:profile-presence-years}
\end{figure*}

\tabref{profile-presence} summarizes the presence of build profiles across all \maven{} packages. \checkNum{45.5\%} of packages contain no build profiles at all, neither in their own POM nor in their parents' POM. \checkNum{12.9\%} define profiles directly in their own POM files, while \checkNum{41.6\%} rely on profiles inherited from parent packages. Overall, \checkNum{54.5\%} of packages include at least one build profile, either declared locally or inherited from a parent POM.
\figref{profile-presence-years} shows the percentage of packages that utilize profiles over the years. We observe a consistent usage of profiles throughout the years in \central{}. Although there have been some fluctuations in the number of \maven{} packages using profiles in the past, in recent years the proportion has remained mostly stable between \checkNum{50\%} and \checkNum{60\%}.

On average, each package contains \checkNum{2.64} build profiles across all levels, with only \checkNum{0.22} profiles defined directly in its own POM. These averages are calculated over the entire \maven{} population. Given that nearly half of all packages do not define any profiles, and only \checkNum{12.6\%} define profiles in their own POMs, we can infer that it is very common for a parent POM to define multiple profiles that are then inherited by its child packages. This suggests that profile usage is widespread in \maven{} projects, and that inheritance plays an important role in how build profiles are structured.

We found \checkNum{345,186} unique build profiles in our dataset. Of these, \checkNum{158,730 (\checkNum{45.98\%})} contain an activation block, while the remaining \checkNum{54.02\%} specify no activation condition and must be activated explicitly. Notably, we identified \checkNum{22,318 (14.1\%)} profiles with JDK-based activation, \checkNum{6,753 (4.2\%)} with OS-based activation, and \checkNum{21,404 (13.5\%)} that are active by default.

\begin{table}
\centering
\caption{Top 5 Profile IDs by Number of Packages}
\label{tab:top5-profile-ids}
\input{tables/top5-profile-ids}
\end{table}

\tabref{top5-profile-ids} lists the five most frequently used profile IDs, based on the number of packages in which they appear, either directly or through inheritance. The ID \code{release} is also the most common, followed by a wide variety of project-specific IDs.
While conducting these analyses, we found evidence that release profiles may impact the reproducibility of packages. For example, in one package~\cite{jsign-parent-4.2}, a profile named \code{release} configures the \maven{} Source Plugin~\cite{mavenSourcePlugin} to generate source JARs specifically for deployment. If this profile is not activated during the release process, these artifacts are omitted. This demonstrates how the activation (or omission) of profiles can alter the set of generated files and, consequently, affect reproducibility.

We also identified \checkNum{134 packages} that define profiles whose ID includes the term \texttt{reproducible}. The most frequent name is reproducible-build, used in \checkNum{121} packages. These profiles often contain configurations aimed at improving build determinism, such as removing non-deterministic metadata. For instance, in one package~\cite{buildsupport-27}, a profile named buildsupport-reproducible is activated only when the \texttt{reproducible} system property is set. This profile enforces reproducibility by specifying a fixed \texttt{outputTimestamp} and configuring plugins to avoid embedding non-deterministic metadata. This example illustrates how profiles can be explicitly designed to strengthen reproducibility by controlling build-time sources of variation.

While investigating the use of release-activated profiles through an analysis of the \maven{} Release Plugin, we found that \checkNum{23,039 (4.87\%)} packages use this plugin and declare one or more profiles to be activated during the release process. The profile ID \code{release} is again the most common in this context. For the sake of brevity, we do not present the full details of the most frequently used IDs in the \maven{} Release Plugin. These are available in our replication package~\cite{repPackage}.

Although the overall use of \maven{} Release Plugin is relatively limited, in the packages that do employ it, these profiles are important for releasing artifacts to \central{}. Identifying them, therefore, provides a useful heuristic for build reproduction.

We found that RC specifies at least one profile using the \code{-P} flag in \checkNum{74.0\%} of the projects it tries to reproduce. We compared the profiles we extracted from the \maven{} Release Plugin with those used by RC. For \checkNum{70\%} of the projects, RC's profiles were identical to those we extracted from the \maven{} Release Plugin. For another \checkNum{6\%} of the projects, the profile we extracted was among those used by RC. In these cases, RC specified multiple profiles, one of which matched ours. Upon closer inspection of the remaining cases, we observed that RC often used a default profile that was not explicitly defined in the project's POM file, and for a few projects, RC used profiles whose origins we could not verify. Overall, our findings indicate that the \maven{} Release Plugin is a reliable source for automatically extracting profiles for most projects.

}

\extension{
\subsection{What Build Tools Do Maven Libraries Use?}
\label{sec:build-tool-extraction}

Studying the build tools used on \central{} is not only essential for automatic reproduction but also provides valuable insights into the technologies and practices that shape the \java{} ecosystem, such as how legacy and modern build systems coexist.

\paragraph{Methodology}
After identifying the build tool as described at the beginning of this section, we compute basic statistics for each build tool.

\paragraph{Results}

For \checkNum{252,654} packages, we identified the build tool using the first approach, by analyzing the metadata inside the compiled artifact. For another \checkNum{66,883} packages, we determined the build tool by inspecting the files in the project repository. Among the successfully classified packages, \maven{} was the most prevalent build tool, accounting for \checkNum{280,016} (\checkNum{87.6\%}) of the packages, followed by \gradle{} with \checkNum{32,806} (\checkNum{10\%}), and \ant{} with \checkNum{6,715} (\checkNum{2\%}).

\begin{figure*}
    \centering
    \includegraphics[width=.9\textwidth]{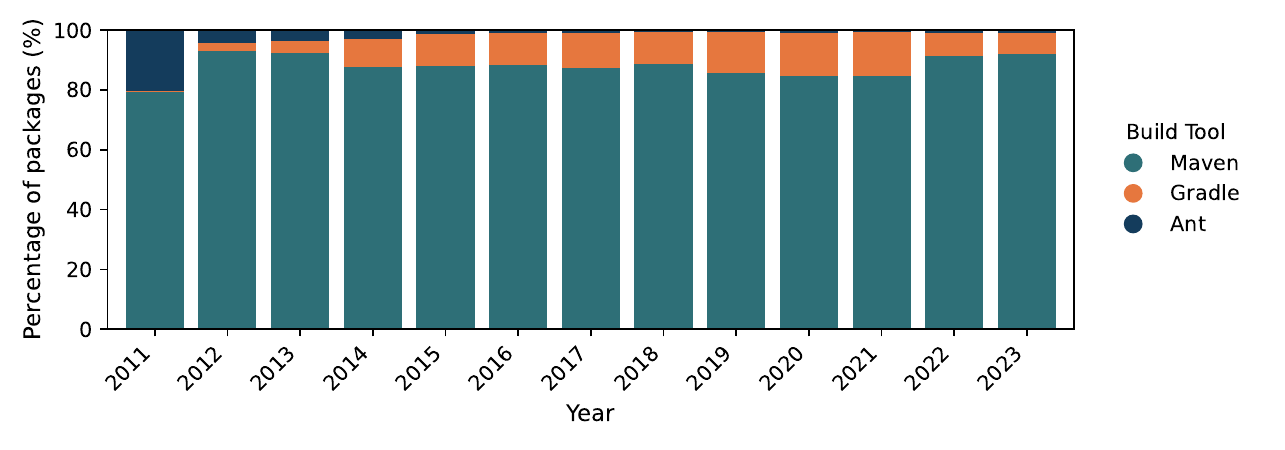}
    \caption{Yearly distribution of packages by build tool. 
    }
    \label{fig:build-tool-years}
\end{figure*}

\figref{build-tool-years} summarizes the yearly distribution of build tools between 2011 and 2023. \maven{} remains dominant across all years, while \ant{}'s popularity has declined in recent years (from more than \checkNum{3,000} packages in 2011 to fewer than \checkNum{300} packages in 2023). In contrast, \gradle{} shows continuous growth, rising from only 47 packages in 2011 to over 5,000 by 2021. These results highlight a clear transition from older, script-based tools like \ant{} to more flexible and modern build systems such as \gradle{}.

RC only includes projects built with \maven{} or \gradle{}, but not with \ant{}. For \gradle{} and \maven{} packages, our identified build tools aligned with RC's results in \checkNum{99.8\%} of the cases.
The only mismatch was a single project that we identified as \gradle{}, while RC classified it as \maven{}. Upon manual inspection, we found that the developers of the package had incorrectly specified the link to the parent project in their POM file. The parent project uses \gradle{}, whereas the project itself is built with \maven{}. As a result, our automated approach detected the build tool of the parent instead of the child. However, when we ran our script on the child repository, it correctly identified \maven{} as the build tool. Therefore, fixing the link in the POM file would automatically resolve this issue.
In this experiment, we also observed a few projects that used multiple build tools. We categorized these projects as \maven{}, consistent with RC. These cases highlight the importance of establishing reasonable default assumptions when designing automated heuristics. Since \maven{} is the most popular and widely supported build tool, especially within RC, it is reasonable for heuristics like ours to prioritize \maven{} during detection or use it as the default option.

\begin{tcolorbox}
\RQ{2}: Can the original build environment be reconstructed after the fact? We automatically extracted the JDK version, line endings, release profiles, and build tools of each release after the fact. When compared with RC, our extracted data showed a high degree of agreement: the JDK version matched in \checkNum{99.5\%} of the cases, line endings in \checkNum{98\%}, release profiles in \checkNum{70\%}, and build tools in \checkNum{99.8\%}.
\end{tcolorbox}
}
\section{\RQ{3}: To What Extent Is It Possible To Automatically Reproduce Maven Artifacts?}
\label{sec:file-level-comparison}

The final research question investigates the overarching research question of this work and analyzes whether an automated approach for reproduction can be successful.
\extension{Out of the \checkNum{479K} packages, \checkNum{235K (49.0\%)} have verified links and recovered tags, and for \checkNum{154K (65.5\%)} of these, we automatically extracted their build tool. To address this research question, we focus exclusively on building these packages. We investigate this research question through multiple sub-questions, each focusing on a specific aspect of reproducibility. For each subquestion, we select a subset of our data suitable for answering it. The details of our data selection are described separately in the corresponding subquestion sections.

}

After building the libraries, we record the build's success status and store the compiler output. We then compare this output to the corresponding release on \central{}. This comparison is facilitated by the RC script. Specifically, it compares the built files with a reference, in this case, the files released on \central{}. While it outputs various data points, we focus primarily on two pieces of information: the \code{ok\_files}, which represents files that match the reference, and the \code{ko\_files}, indicating files that diverge from the reference. Given that RC builds contain entire repositories, a direct comparison of all generated files would not yield meaningful insights. This is because the package under examination might be just a sub-module of the larger repository. A sub-module build generates fewer files compared to the building of the full project. However, as highlighted in \secref{command}, while building the sub-module we also build its associated dependencies. This results in a more extensive file set than what was released for this specific package on \central{}, as the set also contains dependency files. To fix this, we curate a set of files directly related to the \maven{} package in question and limit our comparison to these files. To collect this set, we send a manual request to \central{} and parse the \code{html} list of all files linked to the specific package. This list includes files such as JARs, sources, docs, and POM files.

Reproducibility is a binary concept. An executable is either reproducible, resulting in an identical checksum, or it is not, generating a different checksum. Sometimes, reproducibility is compromised by trivial factors like a timestamp in one file. At other times, it can be due to differing content in multiple class files. Perfect reproducibility facilitates automatic checking but builds often are not reproducible. Thus, determining why a certain executable is not reproducible becomes valuable.
Tools like \tool{diffoscope}~\cite{diffoscope} exist that can compare archives \emph{line-by-line}, but this is too fine-grained for an ecosystem-scale analysis.
As such, we focus on analyzing archive contents \emph{file-by-file} to provide a broader perspective on causes of non-reproducibility.
Given our automatic rebuild, having an automated way to understand how close we are to achieving a perfectly reproducible build is helpful. To achieve this, we compute the MD5 checksum for each generated file and compare these with the corresponding checksums in the reference archive to pinpoint the differences.

\extension{In the first four subquestions of this section, we investigate the effectiveness of \tool{AROMA+} in reproducing packages built by \maven{}, the most popular build tool on \central{} (\checkNum{88\%} of the ecosystem), across four different scenarios. First, we reproduce the same packages that RC previously reproduced to understand how \tool{AROMA+} compares with the existing approach. Next, we apply \tool{AROMA+} to packages not listed in RC to examine whether it can automatically discover new reproducible packages. These packages must include the \code{outputTimeStamp} field to ensure that they at least attempt reproducibility. We then apply \tool{AROMA+} to packages that lack the \code{outputTimeStamp} field to explore whether it can automatically identify near-miss cases. In all these scenarios, we use \tool{AROMA+} without activating any profiles explicitly. As discussed in \secref{rq2}, build profiles are not an official attribute of \code{.buildspec} files and are sometimes specified within the build command. To assess the importance of release profiles for reproducibility, we investigate their impact in a dedicated subquestion. Finally, in the last two subquestions, we evaluate \tool{AROMA+} on the reproducibility of \gradle{} projects, the next most popular build tool on \central{} (\checkNum{10\%} of the ecosystem).}

\subsection{Can AROMA+ Reproduce Reproducible Packages Built by Maven?}
\label{sec:RQ3-1}
\tool{AROMA+} is a tool that automatically extracts build information for a given package. It starts in \central{}, follows repository links to re-establish a source connection, and extracts the necessary information to generate a \code{.buildspec} file.

The previous research questions have shown that we can replicate the \code{.buildspec} files of RC with a high success rate.
In this experiment, we will now investigate how our approach compares to RC on an end-to-end basis.
This is important to investigate because the main purpose of all previous steps is to automate the rebuild process that RC is conducting manually.

\paragraph{Methodology}
We rebuild packages using both \tool{AROMA+} and RC's \code{.buildspec} files, then compare the results.
We select \checkNum{100} random packages built by \maven{} for which RC provides a \code{.buildspec}. First, we run the RC script to reproduce these packages using RC's provided \code{.buildspec} file. Subsequently, we execute the RC script using the \code{.buildspec} generated by our approach. Finally, we compare the number of packages that each method successfully reproduces.

\paragraph{Results}
Out of the \checkNum{100} packages we examined in this experiment, our approach successfully built all of them. However, using RC's \code{buildspecs}, we only managed to build \checkNum{96} packages. A manual inspection revealed that one of these \code{buildspecs} contained a broken repository URL, resulting in a \code{404} error. We noticed that in the RC repository for the newer versions of this package, the link had been updated to match the one we found, but all older versions were still using the broken link. We speculate that the library maintainers changed their repository at some point, and the RC maintainers overlooked updating the links for earlier versions. Nevertheless, when our build was compared to the \central{} release of this specific package, all files matched. Since we could not build this package using RC's approach, we compared the \maven{} files with the \code{ok\_files} present in the RC repository for this package. We found one POM file to be missing. In the other \checkNum{three} instances where the RC build failed, we noted that the command section of the \code{buildspec} initiated with a \code{SHELL} instruction. This command enforces manual intervention, as it prompts an interactive build process for the packages. Consequently, we excluded these from our experiment since we could not automatically build them. However, the build information we extracted was identical to what was present in RC's \code{buildspecs} for these packages.

Out of the remaining \checkNum{96} packages, our approach achieved full reproducibility for \checkNum{10} of them. Similarly, RC reproduced these packages in their entirety. For \checkNum{37} packages, we managed to reproduce all released files except for the \code{sources.jar} files. In contrast, RC was able to reproduce these \code{sources.jar} files. Upon manual inspection, it became clear that the only difference between our build and the RC build was the \emph{release profiles} specified in the command. These \code{37} packages would generate the \code{sources.jar} only when built using a specific release profile. \extension{From a security standpoint, the reproducibility of binaries is the most important goal, and both approaches succeed in this regard. However, to achieve full reproducibility, including all associated artifacts, the role of release profiles is important (See \secref{RQ3-profiles} for detailed profile investigation).} Regarding the remaining \checkNum{49} packages:

\begin{itemize}
\item In \checkNum{11} instances, both our approach and RC reproduced all but the \code{sources.jar} files.

\item For \checkNum{six} packages, neither our method nor RC was able to reproduce the released files. In \checkNum{two} of these cases, the build failed for both methods: one due to a \code{java.lang.ClassNotFoundException} and the other due to a dependency issue.

\item In our build, \checkNum{two} of the packages could not be reproduced, though RC managed to succeed. The primary difference between our method and RC's for these packages was the \emph{release profile}. These two packages specifically required a certain \emph{release profile} for their build.

\item For the remaining \checkNum{30} packages, both methods achieved only partial reproduction. They reproduced some of the released files on \maven{}, but not all, and this time, the unreproduced files extended beyond just the \code{sources.jar}. We acknowledge that anything less than perfect reproducibility in binaries could potentially open the door to attacks. Our intention in discussing partial reproducibility is not to settle for less than perfect. Rather, it is to measure how far non-reproducible packages are from achieving perfect reproducibility. This approach helps us understand the underlying reasons for non-reproducibility and can inform future research about the most crucial missing parts.

\end{itemize}

Overall, our approach achieved very close performance compared to RC when reproducing the packages, despite our method being fully automated. In this experiment, we did not observe any cases where our approach had an inherent limitation.

\subsection{Can AROMA+ Identify New Reproducible Packages Built by Maven That Are Not Listed on RC?}
\label{sec:new-packages-not-in-rc}
Manually creating and maintaining a list of reproducible packages can be challenging and prone to errors.
One potential application of our approach is to assist RC in expanding and updating this list. In this experiment, we devise a scenario to identify reproducible packages not yet documented by RC.

\paragraph{Methodology}
As noted by RC's records, packages that aim for reproducibility do not always fully succeed, often, they are only partially reproducible.
We explore these cases by randomly selecting \checkNum{100} packages that have the \code{project.build.outputTimestamp} property in their POM file, indicating their attempt to be reproducible. We ensure that these packages do not exist in the current RC list.
We then try to recover a \code{.buildspec} file and reuse the RC tooling to attempt reproduction.

\paragraph{Results}
We achieved \checkNum{100\%} reproducibility for \checkNum{five} packages. However, \checkNum{44} of the packages failed to build. This highlights the challenges of rebuilding projects in the wild and highlights the value of automation in deploying reproducible packages and rebuilding them. For \checkNum{23} packages, we managed to reproduce all files except the \code{source.jar}s. For \checkNum{three} packages, even though they were successfully built, we could not reproduce the files. Upon manual inspection, we discovered that the underlying reason was that these packages utilized \emph{release profiles} for their releases on \central{} \extension{(See \secref{RQ3-profiles} for profile investigation).}

For the remaining \checkNum{25} packages, no other files apart from the POM matched the reference.

Overall, the objective of this experiment was to determine whether our approach could be used to expand the RC's list of reproducible \maven{} packages. We pinpointed several packages suitable for inclusion: some with full reproducibility and others with partial reproducibility. As a result, we contributed a few of these packages to the RC repository through \emph{pull requests}, and these were quickly accepted by the main developers of the RC.

\subsection{Find Candidates: Is There Any Near Miss Among Packages Not Tried Reproducibility?}

As discussed in \secref{introduction}, the efforts made so far cover only a limited part of the entire \maven{} ecosystem. Given this context, automated methods that assist in making existing packages reproducible are important. One application of our approach is to detect opportunities for straightforward reproducibility fixes. Addressing such cases, which can be made reproducible with minimal effort, contributes positively to the overall health and security of the ecosystem.

\paragraph{Methodology}
We select \checkNum{100} random packages lacking the \code{project.build.output\-Timestamp} property in their POM file, indicating they have not pursued reproducibility. After building these packages, we conduct a \emph{file-level comparison}, as elaborated in \secref{file-level-comparison}, to identify the reasons for non-reproducibility and to search for near-miss that might be fixed with a small manual effort.

\paragraph{Results}
While the build of \checkNum{39} failed, we were surprised that we managed to partially reproduce some of the other packages, even though they were not designed for reproducibility.
For \checkNum{eight} of these packages, we successfully reproduced the POM files and almost the binary files. The only difference between the binaries and the references was file metadata, like the file dates in the archive itself, suggesting that the content of these packages was otherwise identical. For another \checkNum{26} packages, the only culprits for unreproducibility were the \code{Manifest} and/or \code{pom.properties} files. This indicates that the majority of the content in these packages aligned with what is available on \central{}, and it is the files generated by build tools and plugins during the build process that render them unreproducible. For the remaining \checkNum{27} packages, the unreproducible files also encompassed other types, primarily \code{.class} files. However, the prevalence of this was relatively limited. The package with the most unreproducible class files had \checkNum{43} unreproducible class files out of a total of \checkNum{86}. The subsequent cases had ratios of \checkNum{19 to 53} and \checkNum{13 to 121}, respectively. Apart from these exceptions, all other packages contained fewer than \checkNum{five} unreproducible class files. We also observed an occasional presence of other types of unreproducible files, like XML, but these were largely specific to individual projects and did not form a recurring pattern.

We ignored the reproduction of source files and focused primarily on binaries, which are the most critical aspect of reproducibility.
Sources are available through the recovered repository links.
As we noted, numerous near-miss cases emerged during the experiment. Even though these cases with slight differences might currently not produce the same checksums across different builds, addressing this issue is straightforward. As a result, providing automated recommendations for reproducible releases is feasible in such scenarios. For instance, stripping the timestamp~\cite{ReproducibleBuildMavenPlugin} or assigning it a certain value can make a package reproducible. To demonstrate this, we built a package where the timestamp was the only factor causing unreproducibility. Since \maven{} is immutable (a release cannot be changed after deployment), by manually synchronizing the system's time with that of the released package, we built the package with an identical checksum. Though this technique might appear cumbersome, it highlights the feasibility of automatically repairing the unreproducibility in many packages. This experiment reveals that although the situation on \maven{} may seem bad at first compared to, for example, \tool{Debian}, it can be quickly mitigated. Many packages were \emph{almost} reproducible, but achieving high reproducibility requires the attention of the community.

\extension{
\subsection{How Do Maven Build Profiles Impact Reproducibility?}
\label{sec:RQ3-profiles}

Build profiles in \maven{} allow developers to define conditional build configurations. As discussed in \secref{RQ2-profiles}, these profiles can influence the artifacts produced during the build. In this subsection, we use \tool{AROMA+} to investigate how activating profiles, particularly those referenced by the \maven{} Release Plugin, can impact build reproducibility.

\paragraph{Methodology}

We select \checkNum{100} random packages that (i) declare the \code{project.build\-.outputTimestamp} property in their POM file, (ii) use the \maven{} Release Plugin, and (iii) reference at least one profile through the \code{releaseProfiles} configuration.

We then use RC's script to perform two builds per package: one without manually activating any profiles, and one with all the referenced release profiles explicitly activated in the build command. All other build parameters remain the same and are extracted by \tool{AROMA+}.

After both builds, we compare the outputs. For each package, we compare the list of OK files (files that were bitwise reproducible) between the two builds to determine whether activating profiles improves reproducibility.

\paragraph{Results}

Out of the \checkNum{100} packages, \checkNum{32} showed improved reproducibility when release profiles were activated. The remaining \checkNum{68} packages exhibited no change in the set of OK files. In no case did activating release profiles reduce reproducibility.

For \checkNum{27} of the \checkNum{32} packages, the only additional reproducible artifacts were the \code{sources.jar} files. Therefore, the improved reproducibility stemmed from the successful generation of \code{sources.jar} files, which were missing in the build without a release profile. Our manual investigation of these cases confirms that these artifacts were produced by plugins bound to the release profile, such as the \maven{} Source Plugin, as explained in \secref{RQ2-profiles}. This shows that generating the \code{sources.jar} file only for the final release is a common practice in \maven{}.

In the remaining \checkNum{5} packages, release profiles had a broader impact: in addition to the \code{sources.jar} files, other artifacts, such as additional \code{.jar} files, were also affected. Manual inspection of these cases revealed several observations. In some projects, we observed that reproducibility was not directly the result of using our extracted profile but rather a side effect of it. Some projects use default profiles that serve different purposes but still affect the build output. In some cases, they may even prevent the reproduction of certain files. In these projects, we found evidence that the default profiles were deactivated during the original release. Default profiles are automatically deactivated when any profile is explicitly activated in the build command~\cite{ActivebyDefaultProfiles}. Therefore, we postulate that during the original build, developers used explicit profiles in their commands, which caused the default profiles to be disabled. Consequently, activating the extracted profile on our side also deactivates the default profiles, resulting in the reproduction of additional files that would not be reproduced when no profile is specified. In other cases, we observed the inclusion of project-specific auxiliary files, similar to how the \code{sources.jar} is included only in the final release. Another scenario involved differences in the generated POM file, where build profiles influenced how the released POM was produced, for example, by activating the Flatten \maven{} Plugin~\cite{MavenFlattenPlugin} or modifying certain property values.

These findings highlight that for the majority of projects, release profiles are either unnecessary or only required for reproducing the \code{sources.jar}. Therefore, automation of reproducibility in the \maven{} ecosystem can be achieved to a large extent without release profiles. However, a small fraction of projects cannot be reproduced without them.

}

\extension{
\subsection{Can AROMA+ Reproduce Reproducible Packages Built by Gradle?}

As shown in \secref{build-tool-extraction}, \gradle{} is by far the most commonly used build tool after \maven{} on \central{}. In this experiment, we investigate whether packages built with \gradle{} can be automatically reproduced using \tool{AROMA+}, similar to \maven{} packages.

Although there are artifacts on \central{} built with tools other than \maven{} and \gradle{}, the RC project currently provides automation scripts only for these two. Moreover, as shown in \secref{build-tool-extraction}, \maven{} and \gradle{} together account for \checkNum{98\%} of the ecosystem. Consequently, we limit our experiments to artifacts built with either of these tools.

\paragraph{Methodology}

We randomly select \checkNum{100} packages built with \gradle{} that are already listed in the RC project. We first build these packages using the \code{.buildspec} files from the RC repository, and then use \tool{AROMA+} to generate the \code{.buildspec} files and run RC's build scripts. Finally, we compare the results of both approaches.

\paragraph{Results}

Out of the \checkNum{100} packages that we examined, our approach successfully built all of them, the same as using RC's \code{.buildspec} file. The results of our approach and RC's \code{.buildspec} file were fully identical. We achieved full reproducibility on 18 packages. For 12 packages, we managed to reproduce all files except the \code{sources.jar} files. From the remaining \checkNum{70} packages, \checkNum{20} of the packages had no other file than POM that matched the reference, and the rest achieved partial reproducibility.

\subsection{Can AROMA+ Identify New Reproducible Packages Built by Gradle That Are Not Listed on RC?}

As discussed in \secref{new-packages-not-in-rc}, one potential application of \tool{AROMA+} is to facilitate the expansion of the list of reproducible packages maintained by RC. Reproducing artifacts built with \gradle{} can effectively contribute to this goal. In this experiment, we evaluate the reproducibility of new \gradle{} packages that are not currently listed in RC.

\paragraph{Methodology}

In this experiment, we randomly select \checkNum{100} \gradle{}-built packages that are not formerly listed in RC and that contain the attributes \code{preserve\-FileTimestamps} or \code{reproducibleFileOrder}, which indicate an attempt to be reproducible (see \secref{reproducibility}). We then create a \code{.buildspec} file using the extracted data for these packages. Finally, we run the RC build scripts to attempt reproduction.

\paragraph{Results}
Out of the \checkNum{100} packages, AROMA+ successfully built \checkNum{55}. We achieved full reproducibility for \checkNum{7} of these packages. For an additional \checkNum{16} packages, we successfully reproduced all files except the \code{sources.jar} files. Among the remaining \checkNum{32} packages, \checkNum{21} contained only the POM file that matched the reference. The rest of the packages showed partial reproducibility, where some files aligned with the reference while others did not.

As in \secref{new-packages-not-in-rc}, the objective of this experiment was to evaluate whether our approach could expand the RC's list of reproducible packages, but this time for \gradle{} projects. We identified several \gradle{} packages suitable for inclusion, some fully reproducible and others only partially reproducible.
}

\begin{tcolorbox}
\RQ{3}: To what extent is it possible to automatically reproduce \maven{} artifacts? Overall, automation was possible for \checkNum{32\%} of the packages, of which full reproducibility was achievable for \checkNum{12\%} and partial reproducibility for \checkNum{39\%}.
\end{tcolorbox}

\section{Discussion}
\label{sec:discussion}

The experiments in this paper have touched upon a wide range of areas and resulted in various insights.
We use the discussion section to reflect on actionable insights or recommendations for future work.

\paragraph{Awareness}
Enabling the \code{project.build.outputTimestamp} property in a \maven{} project increases the chance of automated reproducibility substantially.
The only explanation for why this property is not widely used is a lack of awareness among developers.
We postulate that \maven{} tools should start to print a warning about a missing definition during a build (similar to a missing definition of the encoding).
As it does not hurt to define the property, \maven{} could even go one step further and set it automatically, when the property has not been set by developers.
\extension{
Likewise, in \gradle{}, reproducibility can be improved by automatically setting \code{preserveFileTimestamps = false} and \code{reproducibleFileOrder = true} in the build settings.
}

Reproducibility should become a quality attribute of a package. \maven{} could begin warning about non-reproducible dependencies. Once library users start considering reproducibility as a factor in their library selection, library maintainers will have an incentive to put more effort into reproducibility. In the best case, providing \code{.buildspec} files could become a standard in the community.
\extension{
This could be further complemented by richer build metadata. For example, build tools could record details such as the exact command executed and the list of activated profiles included in the final package.
}

On a related note, we found many different release tagging styles in our analysis.
Developers need to realize that standardization facilitates automated processing and enhances the transparency of the ecosystem.
Instead of reinventing the wheel, we recommend that developers adhere to the most popular community conventions for tagging their releases.

\paragraph{Better Research}
We currently observe that the absence of \maven{} source archi\-ves (\code{sources.jar}) often leads to the exclusion of packages from studies that require source code analysis. For instance, in the study by Karakoidas et al.~\cite{Karakoidas2015}. Recovering repositories and release tags for \maven{} packages not only enhances the likelihood of better reproducibility but can also have a positive impact on these research efforts.

\paragraph{Sorry State of \central{}}
We find it worrisome to observe how many projects cannot be reproduced simply because the necessary resources are not available or are no longer accessible. Future work should focus on investigating the feasibility of removing broken artifacts from the repository to restore \central{} to a fully self-contained and reproducible state. Broken packages no longer serve a purpose and impede maintenance efforts.

We believe that the strong separation between hosting binaries and source code, as designed in \central{}, is the most likely explanation for this state. Perhaps the concept of a central repository is outdated and requires re-evaluation. Current practices on major hosting platforms like \tool{GitHub} or \tool{GitLab} highlight the distributed nature of \maven{} and bring sources and releases closer together in their integrated package registries. However, it is important to note that a fully decentralized system may emphasize the problems with data consistency that we have identified on \central{}, where, at least in theory, regulations exist to maintain a basic level of repository hygiene.

\paragraph{Relevance} Our experiments showed that automated reproduction could potentially be attempted for 154K out of 479K packages (32\%), and out of the packages we attempted, only 72 out of 600 (12\%) were fully reproducible, with 236 out of 600 (39\%) being at least partially reproducible. While these numbers may seem small, they represent automated results that allow us to focus manual efforts on more challenging packages.
If the same fractions were to apply to the 10M existing \maven{} packages, we would be able to automatically reproduce approximately \checkNum{$32\% \times (12\% + 39\%) \times 10M \simeq 1.6M$} packages, significantly expanding the RC dataset beyond the existing 9K packages. We believe that \tool{AROMA+} has significant potential, and there are ample opportunities for further improvement in future research.

\extension{
\paragraph{Future Directions} Currently, reproducibility tooling for \maven{} compares the checksums of files built locally with those published on \central{}. It does not detect files that exist on \central{} but are missing from the local build. If a local file has the same checksum, it is categorized as \code{ok}; if the checksum differs, it is marked as \code{ko}. We argue that a third category, \code{missing}, should be introduced.
\central{} is a critical hub on which many projects depend, and establishing a high level of trust by verifying all hosted content is therefore essential. It would be detrimental if malicious content were introduced into \central{} through the addition of new files rather than the modification of existing ones, as such cases could go unnoticed under the current tooling. Future work should investigate the impact of this limitation and explore ways to address it.
}

\extension{
In this study, we showed that release profiles are typically required to reproduce \code{sources.jar} files. While the main binaries, as the most security-critical components, can be reproduced in most cases without profiles, in a few instances, profiles were also needed to reproduce them. Therefore, although achieving reasonable overall reproducibility for \central{} may not require prioritizing release profiles, achieving full reproducibility across the entire ecosystem does necessitate their automatic extraction. In this study, we identified the required build profiles by analyzing the configuration of the \maven{} Release Plugin~\cite{mavenReleasePlugin} and evaluated the impact of activating these profiles on reproducibility. However, profile activation mechanisms are highly complex and remain insufficiently explored. Future research could develop more diverse and advanced heuristics to automatically detect essential profiles. This line of work could also be extended to similar conditional configurations in other JVM build tools, such as \gradle{}.
}

\extension{In this study, we also observed the existence of multi–build-tool projects. A promising future direction is to investigate such projects by identifying the coexistence of different tools and comparing their impact on reproducibility.}

Another future direction could be to formulate reproduction as a search problem. The goal is to prune impossible environmental variables and perform a grid search to find a valid \code{.buildspec}.
Even cases that are not or only partially reproducible right now can inform such an automated approach and support the manual creation of a \code{.buildspec}.
Future work should extend \tool{AROMA+} and explore the localization and repair of unreproducible builds, especially cases in which large parts of the release could be matched.
This would be invaluable for the ecosystem and could enhance its security.

\subsection{Threats to Validity}
The following limitations should be taken into account when interpreting the results. 

\paragraph{Internal Validity}
We conducted manual inspections and implemented heuristics in this research, which might introduce human errors or overlook corner cases. To minimize potential issues, we carefully documented all of our steps. Furthermore, for the manual inspections, two authors were involved to mitigate the chances of mistakes. We also compared our results to a manually curated baseline to estimate precision at each step.
We reused existing open-source tools for rebuilding packages, which are widely accepted and integrated with \maven{}.
However, there is a possibility of inheriting any flaws that they might have. To mitigate this, we built a flexible infrastructure for others to integrate their tools and compare their results with ours. We also acknowledge that hidden bugs in our implementation could exist. To address this, we conducted multiple manual tests, created an extensive test suite, and made our data and code public for community review. 

\paragraph{External Validity}
In our study, while implementing the heuristics, we focused on common patterns. For example, our study was limited to git-based packages, excluding other version control systems like \tool{Subversion} and \tool{Mercurial}. We believe that this is only a small limitation that mainly affects releases that are older than a decade, as documented in our analysis of the popularity of hosting platforms.
We encourage future work to replicate our study in other ecosystems like \code{NPM} or \code{PyPI}.

\paragraph{Construct Validity}
When extracting the \tool{JDK} version for cases where it was not present in the manifest file, we utilized the \java{} version specified in the POM. While this version is not necessarily guaranteed to align precisely with the \tool{JDK} used to build the project, it typically serves as a lower-bound approximation. This approach might not provide the maximum precision, but it enables us to build the packages using various versions and then compare the results. This allows for selecting the most probable version that was used in the library release.

\section{Related Work}
\label{sec:related-work}

\paragraph{Software Ecosystems}
Multiple studies explored various aspects of software ecosystems. Kula et al.~\cite{Kula2015Trusting,Kula2017Exploratory} investigated software updates and library life cycles in \maven{}. Bavota et al.~\cite{Bavota2015Apache} examined changes within the Apache ecosystem, exploring elements such as dependency graphs, project size, releases, and client behavior during dependency version upgrades. D\"{u}sing et al.~\cite{dusing2022analyzing} researched library updates for patching vulnerabilities and found that a considerable number of \maven{} vulnerabilities were patched before disclosure. Mir et al.~\cite{mir2023effect} studied the propagation of vulnerabilities in the \maven{} ecosystem.
Raemaekers et al.~\cite{Raemaekers2013} developed the \maven{} Dependency Dataset, which promoted research into the software evolution of the \maven{} Repository. Keshani et al.~\cite{keshani2021scalable} proposed a scalable call graph generation approach for \maven{} and used these call graphs to investigate the correlation between method popularity and breaking changes~\cite{keshani2023relation}. Karakoidas et al.~\cite{Karakoidas2015} collected comprehensive code metrics associated with object-oriented design, package design, and program size in \maven{} libraries. Soto-Valero et al.~\cite{soto2021comprehensive} studied the occurrence of bloated dependencies in the \maven{} ecosystem. Soto-Valero et al.~\cite{soto2019emergence} researched library version usage and distribution in the \maven{} ecosystem. Mitropoulos et al.~\cite{mitropoulos2014bug} investigated the \central{} Repository using the \tool{FindBugs} tool to establish a link between artifact size and bug count.
Abdalkareem et al.~\cite{abdalkareem2017developers} studied trivial packages in the \tool{npm} ecosystem, while Cogo et al.~\cite{cogo2021empirical} analyzed same-day releases in the \tool{npm} ecosystem. Kula et al.~\cite{kula2017modeling} introduced the Software Universe Graph (SUG) for software ecosystems analysis and compared dependency update behavior between \maven{} and \tool{CRAN}. Benelallam et al.~\cite{benelallam2019maven} constructed an artifact-level model of the \central{} Repository to identify duplicated artifacts, while Kanda et al.~\cite{kanda2014measuring} explored the occurrence and duplication of inner JAR files within the \central{} Repository's JAR files.

\paragraph{Software Builds}

Several studies delved into the software builds. Ma et al.~\cite{ma2020empirical} analyzed \maven{} archetypes, identifying prevalent schema patterns in POM files. Zhang et al.~\cite{Zhang2022BuildSonic} presented \tool{BuildSonic}, a tool that pinpoints and repairs configuration issues in both \maven{} and \gradle{} builds, subsequently enhancing build speeds. Tamaraw et al.~\cite{tamrawi2012build} attempted to aid in the maintenance of build script files through static analysis. Keshani et al.~\cite{keshani2024frankenstein} proposed a lightweight call graph generation approach for software builds.
Tufano et al.~\cite{Tufano2017} examined cases of uncompileable snapshots in \java{} projects that use \maven{}, discovering that most projects experienced such snapshots due to dependency resolution issues. Gazzillo et al.~\cite{gazzillo2017kmax} identified all build configurations using their proposed \tool{Kmax}. Sotiropoulos et al.~\cite{sotiropoulos2020model} introduced \tool{BuildFS}, which models build executions and identifies faults in build systems. Hassan et al.~\cite{hassan2018hirebuild} proposed \tool{HireBuild}, a history-driven approach for repairing build scripts. \tool{HoBuff}~\cite{lou2019history} further refined \tool{HireBuild} by taking into account both the current project and external resources. Lastly, Lou et al.~\cite{lou2020understanding} analyzed 1,000 build-related issues on \tool{Stack Overflow} and summarized the fix patterns for three build systems: \maven{}, \ant{}, and \gradle{}.

\paragraph{Reproducibility}
In 1984, Ken Thompson reflected on the \emph{Trusting Trust Attacks}. These attacks involve compilers being compromised to embed malicious Trojan horses into the compiled code~\cite{thompson1984reflections}. Since these alterations are not evident in the source code itself, detecting them is especially challenging. In response to this, Wheeler et al.~\cite{wheeler2005countering} introduced the concept of diverse double compiling. This method compares the compilation results of the same source code using different compilers to prevent malicious attacks. This research paved the way for the initiative known as reproducible builds~\cite{reproducibleBuilds}. This initiative offers a set of practices to enhance the reproducibility of software packages. Several studies sought to expand on these ideas. Holler et al.~\cite{holler2015evaluation} explored diverse double compilations in embedded systems. Shi et al.~\cite{shi2021experience} developed a unified build process along with a toolset for verifiable builds. \tool{RepLoc}~\cite{ren2018automated} and \tool{RepTrace}~\cite{ren2019root}, aimed to pinpoint the origins of reproducibility issues. \tool{RepFix}~\cite{Ren2022Automated} and \tool{ConstBin}~\cite{he2020constbin} work towards automatically fixing unreproducible builds. Another solution, \tool{DetTrace}~\cite{navarro2020reproducible}, introduces a container abstraction for \tool{Linux}, ensuring reproducibility.
Several studies delved into the state of reproducibility and associated challenges within the open-source ecosystem. Fourne et al.~\cite{fourne2023s} explored the motivations, challenges, and solutions related to reproducibility by interviewing developers. Butler et al.~\cite{butler2023business} interviewed business managers to understand the adoption of reproducible builds in businesses, shedding light on the technical reasons for embracing these practices. Lamb et al.~\cite{lamb2021reproducible} discussed what it means to build software reproducibly. Carnavalet and Mannan~\cite{de2014challenges} carried out an empirical study on reproducible builds within security-critical software, summarizing the practical challenges of reproducibility. \extension{ Sharma et al.~\cite{sharma2025causescanonicalizationunreproduciblebuilds} analyzed unreproducible \java{} artifacts and proposed a taxonomy of six root causes of unreproducibility,
including differences in build manifests, SBOMs, filesystems, JVM bytecode,
versioning properties, and timestamps. They further introduced canonicalization as a mitigation strategy.
Dietrich et al.~\cite{dietrich2024levels}\cite{10989044} studied the problem of comparing binaries produced by independent build providers and provided a framework for assessing equivalence between binaries.

Pohl et al.~\cite{pohl2025sok} studied reproducibility within scripting language ecosystems such as \tool{JavaScript}, \tool{Python}, and \tool{Ruby}. They highlighted how these ecosystems face unique challenges due to their dependence on package registries that often lack verifiable links between distributed artifacts and their source code.
Benedetti et al.~\cite{benedetti2025empirical} present a large-scale empirical study of reproducible builds across six major software ecosystems. They show that while reproducibility currently varies widely, nearly all packages can achieve reproducible builds through modest configuration and tooling improvements.
Dietrich et al.~\cite{dietrich2025daleq} introduce a tool that determines and explains the equivalence of non-bitwise-identical \java{} binaries by disassembling them into relational databases.
}

Despite the significant contributions of these studies, to the best of our knowledge, no current research addresses the issue of automatically reproducing \maven{} libraries. Our study aims to fill this gap and offer an in-depth understanding of library reproducibility in \central{}.

\section{Conclusions}\label{sec:conclusions}

While the software development community has made significant progress in enhancing the reproducibility of libraries, there remains a considerable gap in understanding and addressing this issue within the \maven{} ecosystem. The existing focus on \tool{Debian} has left the \maven{} ecosystem relatively understudied, with RC's list of reproducible \maven{} libraries being both limited and challenging to maintain. Our research tried to bridge this gap by automatically finding the reproducible \maven{} packages and their build environments. This enabled us to automatically rebuild the source code of these libraries and compare the results with the files available on \maven{}.
While the automation is still limited to a subset of all releases for which a basic set of features can be recovered, the results of our research are highly promising.
Our experiments demonstrate that we can achieve a \checkNum{99.8\%} similarity to manually crafted \code{.buildspec} files.
Using our approach we found previously missing reproducible libraries and contributed them to the RC repository.
Furthermore, we made our dataset and tools openly accessible to the public~\cite{repPackage}.

\begin{acknowledgements}
We would like to thank the co-authors of the conference version of this paper, Tudor-Gabriel Velican, Gideon Bot, and Sebastian Proksch, who were unfortunately unable to contribute to the journal version.
\end{acknowledgements}

\section{Declarations}

\subsection*{Funding} This research received no external funding.

\subsection*{Ethical approval}
Not applicable. This study did not involve human participants or animals.

\subsection*{Informed consent}
Not applicable. No human subjects were involved in this research.

\subsection*{Author Contributions}
Mehdi Keshani, Amirhossein Rahmati, Mohammad Hossein Aref, and Abbas Heydarnoori contributed to the conception, design, analysis, and writing of the manuscript. All authors read and approved the final version of the paper.

\subsection*{Data Availability Statement}
Our replication package~\cite{repPackage} includes the code, data, and all essential materials needed to reproduce the results presented in this paper.

\subsection*{Conflict of Interest}
The authors declare that they have no conflict of interest.

\subsection*{Clinical Trial Number}
Clinical trial number: not applicable.

\bibliographystyle{spmpsci}
\bibliography{references}

\end{document}

%% file: tables/input-breakdown.tex
\ra{0.9}
\small
\begin{tabular}{@{}lrr@{}}
    \toprule
    \textbf{Set} & \textbf{\#Packages} & {\bf \%} \\ \midrule
    Indexed packages& 10,333,041 & 100.0 \\
    Unique group:artifact combinations& 479,915 & 4.6 \\
    \midrule
    Sample size of study& 479,915 & 100.0 \\
    Successfully downloaded& 473,352 & 98.6 \\
    Have archives& 410,102 & 86.6 \\
    \bottomrule
\end{tabular}

%% file: tables/url-field-usage.tex
\ra{0.9}
\small
\begin{tabular}{@{}lrrcrr@{}}
    \toprule
    & \multicolumn{2}{c}{\bf Found} && \multicolumn{2}{c}{\bf Valid} \\
    \cmidrule{2-3} \cmidrule{5-6}
    {\bf Metric} & \#Packages & \% (of Total) && \#Packages & \% (of Found) \\ 
    \midrule
    Has $\geq$ 1 URL field & 448,273 & 98.6 && 354,948 & 79.2 \\ 
    Has url & 442,379 & 97.3 && 166,742 & 37.7 \\  
    Has scm.url & 442,360 & 97.3 && 327,744 & 74.1 \\ 
    Has scm.connection & 425,999 & 93.7 && 224,769 & 52.8 \\ 
    Has scm.developerConnection & 356,982 & 78.6 && 117,912 & 33.0 \\ 
    \bottomrule
\end{tabular}

%% file: tables/top10-tagging-patterns.tex
\ra{0.9}
\small
\begin{tabular}{@{}lllr@{}}
    \toprule
    {\bf Pattern} & {\bf Example coordinate} & {\bf Tags} & {\bf \%Packages} \\ 
    \midrule
    \code{v\VER} & com.ethlo.dachs:dachs-audit:1.0.0 & v1.0.0 & 48.2 \\
    \code{\VER} & com.eriwen:groovyrtm:2.1.1 & 2.1.1 & 34.9 \\
    \code{artifactId-\VER} & com.expedia.tesla:tesla:4.0 & tesla-4.0 & 6.4 \\
    \code{p1-\VER} & org.activiti:activiti-form-api:6.0.0 & activiti-6.0.0 & 5.3 \\
    \code{release-\VER} & com.senseidb.clue:clue:0.0.2 & release-0.0.2 & 1.5 \\
    \code{p1-p2-\VER} & com.aoapps:ao-dbc-book:3.1.1 & ao-dbc-3.1.1 & 1.2 \\
    \code{release/\VER} & com.github.dalet-oss:vfs-gcs:2.2.0 & release/2.2.0 & 0.4 \\
    \code{p1-p2-p3-\VER} & com.adobe:aio-lib-java-ims:0.0.4 & aio-lib-java-0.0.4 & 0.3 \\
    \code{\VER-release} & com.github.machaval:cli\_2.12:0.2.0 & 0.2.0-release & 0.3 \\
    \code{v.\VER} & dev.struchkov.haiti:haiti-core:1.0.3 & v.1.0.3 & 0.2  \\  
    \bottomrule
\end{tabular}

%% file: tables/profile-presence.tex
\ra{0.9}
\small
\begin{tabular}{@{}lrr@{}}
    \toprule
    \textbf{Profile Presence} & \textbf{Number of Packages} & \textbf{Percentage} \\
    \midrule
    No Profiles & 215{,}437 & 45.5\% \\
    Profiles Only in Parent POMs & 197{,}010 & 41.6\% \\
    Profiles in Own POM & 60{,}844 & 12.9\% \\
    \bottomrule
\end{tabular}

%% file: tables/top5-profile-ids.tex
\ra{0.9}
\small
\begin{tabular}{@{}lr@{}}
    \toprule
    \textbf{Profile ID} & \textbf{Number of Packages} \\
    \midrule
    \texttt{release} & 98{,}402 \\
    \texttt{sonatype-oss-release} & 51{,}934 \\
    \texttt{apache-release} & 25{,}543 \\
    \texttt{release-sign-artifacts} & 22{,}213 \\
    \texttt{doclint-java8-disable} & 18{,}261 \\
    \bottomrule
\end{tabular}